\documentclass[a4paper,12pt]{article}
\usepackage{amssymb}
\usepackage{amsmath}
\usepackage{mathtools}
\usepackage{slashed}
\usepackage{color}
\usepackage{xcolor}
\usepackage{indentfirst}
\usepackage{graphicx}

\usepackage[T1]{fontenc}
\usepackage[utf8x]{inputenc}

\usepackage{csquotes} 
\usepackage{caption}
\usepackage{comment}
\usepackage{cite}
\usepackage{mathrsfs}
\usepackage[unicode]{hyperref}
\usepackage{lscape}
\usepackage{float}

\hypersetup{
    colorlinks=false,
    citebordercolor=lime,
    linkbordercolor=pink,
}

\numberwithin{equation}{section}

\renewcommand{\thefootnote}{\fnsymbol{footnote}}

\begin{document}

\title{
\begin{flushright}
\begin{minipage}{0.2\linewidth}
\normalsize
KEK-TH-2322 \\*[50pt]
\end{minipage}
\end{flushright}
{\Large \bf 
Sharpening the Boundaries Between Flux Landscape and Swampland by Tadpole Charge
\\*[20pt]}}

\author{Keiya Ishiguro$^{a}$\footnote{
E-mail address: ishigu@post.kek.jp
}
\ and\
Hajime~Otsuka$^{b}$\footnote{
E-mail address: hotsuka@post.kek.jp
}\\*[20pt]
$^a${\it \normalsize 
Graduate University for Advanced Studies (Sokendai),}\\
{\it \normalsize 1-1 Oho, Tsukuba, Ibaraki 305-0801, Japan.} \\
$^b${\it \normalsize 
KEK Theory Center, Institute of Particle and Nuclear Studies,}\\
{\it \normalsize 1-1 Oho, Tsukuba, Ibaraki 305-0801, Japan}}
\maketitle

\date{
\centerline{\small \bf Abstract}
\begin{minipage}{0.9\linewidth}
\medskip 
\medskip 
\small
We investigate the vacuum structure of four-dimensional effective theory arising from Type IIB flux compactifications on a 
mirror of the rigid Calabi-Yau threefold, corresponding to a T-dual of the DeWolfe-Giryavets-Kachru-Taylor model in Type IIA flux compactifications. 
By analyzing the vacuum structure of this interesting corner of string landscape, it turns out that there exist perturbatively unstable de Sitter (dS) vacua in addition to supersymmetric and non-supersymmetric anti-de Sitter vacua.
On the other hand, the stable dS vacua appearing in the low-energy effective action violate the 
tadpole cancellation condition, indicating a strong correlation between the existence of dS vacua 
and the flux-induced D3-brane charge (tadpole charge). 
We also find analytically that the tadpole charge constrained by the tadpole cancellation condition emerges in the scalar potential in a nontrivial way.
Thus, the tadpole charge would restrict the existence of stable dS vacua, and this fact underlies the statement of the dS conjecture. 
Furthermore, our analytical and numerical results exhibit that distributions of ${\cal O}(1)$ parameters in expressions of several swampland conjectures peak at specific values. 
\end{minipage}
}

\renewcommand{\thefootnote}{\arabic{footnote}}
\setcounter{footnote}{0}
\thispagestyle{empty}
\clearpage
\addtocounter{page}{-1}

\tableofcontents

\section{Introduction}
\label{sec:1}

String compactification is one of the major topics in string theory if one expects it to be the fundamental theory of our world. 
One way to understand what four-dimensional (4D) fields derived from ten-dimensional (10D) 
superstring theory is the Kaluza-Klein (KK) decomposition on the extra six-dimensional (6D) space. 
Then, zero-modes are supposed to play an important role in 4D spacetime. 
The existence of 4D supersymmetry (SUSY) requires the 6D space to 
equip $SU(3)$ holonomy, and it is proved that Calabi-Yau (CY) threefolds possess this structure. 
Hence, CY threefolds and toroidal orbifolds (considered singular limits of CY threefolds) are of 
particular interest.  
The 4D effective theories arising from string compactifications contain a multitude of massless fields called moduli fields, reflecting the geometry of the corresponding internal manifold. 
These moduli fields appear in 4D Yukawa couplings and gauge couplings such that those free parameters in the Standard Model are determined by their vacuum expectation values (VEVs) in string compactifications. 
Furthermore, the dynamics of moduli fields are closely related to the construction of de Sitter (dS) 
vacua. 
Although the moduli stabilization and string landscape have been studied intensively in the past two decades, there are several developments from the viewpoint of the swampland program \cite{Vafa:2005ui,ArkaniHamed:2006dz,Ooguri:2006in}, stating that the 4D low-energy effective field theories should satisfy several conjectures to admit an ultra-violet completion to a consistent theory of quantum gravity. (See for a review, e.g., Ref. \cite{Palti:2019pca}.)

After the moduli stabilization, the vacuum energy behaves as the cosmological constant $\Lambda$.
With theoretical and phenomenological motivations, one needs to find the vacuum with the positive vacuum energy to realize the accelerating universe. However, it was known that the stable classical de Sitter (dS) vacua are difficult to achieve in string compactifications as stated in 
the Maldacena-Nu$\tilde{\rm n}$ez no-go theorem \cite{Maldacena:2000mw}. (For 
an overview of the construction of dS vacua in the string theory, see, e.g., Ref. \cite{Danielsson:2018ztv}.) 
The absence of stable classical dS vacua motivates us to consider the no-go conjectures of dS vacua, known as the dS swampland conjecture \cite{Obied:2018sgi,Andriot:2018wzk,Garg:2018reu,Ooguri:2018wrx}. 
So far, the dS swampland conjecture is well established in Type IIA flux compactifications \cite{Camara:2005dc,DeWolfe:2005uu}, since all the closed string moduli fields can be stabilized by Ramond-Ramond (RR) and Neveu-Schwarz-Neveu-Schwarz (NS) fluxes. 
Indeed, the dS swampland conjecture is satisfied for some examples of Type IIA flux vacua \cite{Hertzberg:2007wc} and more general classes in a parametrically controlled regime \cite{Junghans:2018gdb}. Consequently, the obtained flux vacua fall Anti-de Sitter (AdS) vacua or dS vacua with tachyonic directions.\footnote{For the recent developments in Type IIA flux compactifications, see, e.g., Refs. \cite{Marchesano:2020uqz, Shukla:2019dqd}}
Those AdS vacua satisfy the AdS/moduli scale separation conjecture \cite{Gautason:2018gln} and the AdS distance conjecture \cite{Lust:2019zwm} in the supersymmetric case as pointed out in Refs. \cite{Danielsson:2018ztv,Lust:2019zwm,Marchesano:2019hfb}. These conjectures prohibit the parametric separation between the mass of towers of the lightest states and the AdS radius. 

In Type IIB flux compactifications, K\"ahler moduli describing the size of the internal manifold cannot be stabilized by RR and NS three-form fluxes as opposed to the axio-dilaton and complex structure moduli. (See for a review, Ref. \cite{Blumenhagen:2006ci}.) 
It is then required non-perturbative effects and/or $\alpha'$-corrections to stabilize the K\"ahler moduli as demonstrated in the  Kachru-Kallosh-Linde-Trivedi scenario \cite{Kachru:2003aw} and large volume scenario \cite{Balasubramanian:2005zx},\footnote{For the perturbative stabilization of K\"ahler moduli, we refer the reader to, e.g., Refs. \cite{Berg:2005yu,Kobayashi:2017zfd}.} 
where the AdS minima are uplifted to the dS vacua by some mechanism. 
The validity of the dS vacua obtained by employing non-perturbative effects and uplifting sources should be justified from several aspects \cite{Gautason:2018gln,Bena:2018fqc,Blumenhagen:2019qcg,Gao:2020xqh}. 
The rigorous check of swampland conjectures is still an important open issue to understand what constitutes exact boundaries between the string landscape and the swampland.

In this paper, we present a different approach in Type IIB flux compactifications to resolve this issue. 
We deal with ``non-geometric'' CY manifold with the Hodge number being $h^{1,1} = 0$,\footnote{Here and in what follows, we use the term ``non-geometric'' in this sense.} 
i.e., no deformation of the K\"ahler structure, and the volume is already fixed by symmetries \cite{Candelas:1993nd}. 
The manifold can be described as a mirror of a certain ``rigid'' CY manifold with $h^{2,1}=0$ but $h^{1,1} \neq 0$. 
The prototypical example of Type IIA flux compactifications called DeWolfe-Giryavets-Kachru-Taylor (DGKT) model \cite{DeWolfe:2005uu} is indeed based on the rigid CY threefold, namely $T^6/(\mathbb{Z}_3\times \mathbb{Z}_3^\prime)$ 
toroidal orbifold. 
In this paper, we examine Type IIB flux compactifications corresponding to a T-dual of the DGKT model.\footnote{See for the Type IIB flux compactification on the rigid $T^6/(\mathbb{Z}_3\times \mathbb{Z}_3^\prime)$ orbifold, Ref.\cite{Shukla:2019akv}.} 
Using the fact that one can stabilize all the closed string moduli by three-form flux themselves and deal with more general fluxes than the T-dual IIA side\footnote{Throughout this paper, we focus on the stabilization of untwisted complex structure moduli.}, it is expected that this setup captures some crucial properties of flux vacua and string landscape. 
So far, this class of manifolds has been studied in Refs. \cite{Becker:2006ks,Becker:2007dn}, but the vacuum structure is not fully understood. 
For this reason, we wish to set this manifold as a testing ground for various swampland conjectures. 
In particular, we focus on the AdS/moduli scale separation conjecture, the AdS distance conjecture, and the dS swampland conjecture.

Our findings of the flux vacua on the mirror of rigid CY are summarized as follows:

\begin{itemize}

    \item We find the SUSY and non-SUSY stable AdS flux vacua\footnote{In this paper, we call non-SUSY minima appearing from the 4D effective field theory (EFT) the non-SUSY vacua. Those minima are reasonable under the assumptions that the string coupling constant $g_S \ll 1$ and the KK mode mass $\gg$ moduli mass from the viewpoint of the 4D EFT. However, we have not analyzed effects of possible corrections including the flux backreaction on the non-SUSY minima. Thus, it does not mean that we have found a fully consistent non-SUSY vacuum in the string compactification.} for a large region of flux quanta satisfying the tadpole cancellation condition.

    \item As discussed in detail in section \ref{sec:vacua structure}, it is found that the tadpole charge is severely constrained by the stability condition to allow the existence of Minkowski and dS vacua in addition to the well-known tadpole cancellation condition. 
    It indicates that the realization of the stable dS vacua would be difficult to be realized in the string landscape due to the small number of tadpole charges constrained by the tadpole cancellation condition.

    \item There exists a new class of dS vacua with tachyonic directions for a limited region of flux quanta taking into account the tadpole cancellation condition\footnote{Note that the existence of tachyon was observed in a broad class of dS vacua in 10D Type II supergravities with $D_p$-branes and orientifold $O_p$-planes and for more details, we refer to, for instance, Ref. \cite{Danielsson:2018ztv}.}, whereas the perturbatively stable dS vacua appearing in the low-energy effective action violate the tadpole cancellation condition\footnote{Hence, the stable dS vacua are not justified even in the 4D EFT. Moreover, it is also possible that there is a true vacuum with lower energy. Our purpose is to reveal the vacuum (minimum in the 4D EFT) structure through the tadpole charge.}. Furthermore, the structure of AdS vacua depends on the tadpole charge as discussed in the AdS/moduli scale separation and the AdS distance conjectures. 
    It indicates that the tadpole charge would determine the boundaries between the string landscape and the swampland. 
    
    \item Distributions of $\mathcal{O}(1)$ parameters arising in expressions of swampland conjectures are 
    investigated for both SUSY and non-SUSY flux vacua. 
    In particular, our analytical and numerical results exhibit that these parameters peak around specific values for the AdS/moduli scale separation conjecture, and a strong version of AdS distance conjecture holds at the SUSY AdS vacua, irrespective of fluxes.
\end{itemize}

This paper is organized as follows: 
In section \ref{sec:fluxcompactification}, we briefly review the flux compactifications on the mirror of rigid CY manifold, corresponding to a specific Landau-Ginzburg (LG) model. 
The complex structure moduli can be identified with deformations of the superpotential in a traditional way. 
Since the background geometry is described by the special geometry on a complex threefold, it enables us to obtain an exact form of the Gukov-Vafa-Witten (GVW) type superpotential \cite{Gukov:1999ya}. 
Then, all the moduli fields can be stabilized in the context of flux compactifications, taking into account the effects of fixed K\"ahler moduli in the K\"ahler potential.  
On the basis of the 4D effective action in section \ref{sec:fluxcompactification}, the vacuum structure is analytically examined with an emphasis on the role of tadpole charge in section \ref{sec:vacua structure}. 
In section \ref{sec:test}, we investigate three Swampland conjectures. 
Finally, we summarize the results of this paper and discuss some remaining issues in section \ref{sec:con}. 

\section{Brief Review of Type IIB Flux Compactifications on the Mirror of Rigid CY}
\label{sec:fluxcompactification}

In section \ref{sec:review of underlying manifold}, we begin with the construction of the mirror of rigid CY threefold with the $1^9/\mathbb{Z}_3$ Gepner model, which allows us to study the Type IIA/IIB flux compactifications in a unified way. 
After briefly reviewing Type IIB flux compactifications with special geometry in section \ref{sec:flux compactification with special geometry}, we derive the effective moduli action in section \ref{sec:effaction}, where the dilaton K\"ahler potential receives the correction from the effect of fixed K\"ahler moduli. 
We refer the details of the background geometry to Refs. \cite{Candelas:1993nd, Becker:2007dn}. 

\subsection{The \texorpdfstring{$\tilde{\mathscr{Z}}$}{Z \unichar{"0303}} manifold}
\label{sec:review of underlying manifold}

Let us consider the $1^9/\mathbb{Z}_3$ Gepner model, where the LG worldsheet superpotential is given by
\begin{align}
  W = \sum_{k=1}^{9} y_k^3, 
  \label{eq:1^9 superpotential}
\end{align}
subject to
\begin{align}
      \mathbb{Z}_3: y_k \rightarrow \omega y_k,
\end{align}
with $\omega = e^{\frac{2\pi i}{3}}$ .
On the vanishing locus of $W$, i.e., $W=0$, there exist $84$ $\mathbb{Z}_3$-invariant polynomials and the background geometry corresponds to $\mathbb{P}_8[3]$ manifold through the LG-CY correspondence. 
Based on this Gepner model, the class of rigid CY $\mathscr{Z}$ is constructed, and correspondingly, the class of mirror CY $\tilde{\mathscr{Z}}$ is provided by a quotient of $\mathbb{P}_8[3]$, namely $\tilde{\mathscr{Z}}=\mathbb{P}_8[3]/G$ 
with a certain group $G$. 
More precisely, these mirror CYs have seven dimensions and belong to a class called the ``generalized CY'' (see for details, Ref. \cite{Candelas:1993nd}).

What is important on this sevenfold is that its middle-cohomology structure is the same as that of usual CY threefolds. Under the Hodge decomposition $H^7(\tilde{\mathscr{Z}}) = \bigoplus_{p+q=7} H^{p, q}_{\bar{\partial}}(\tilde{\mathscr{Z}})$, the corresponding Hodge number is listed as
\begin{align}
    [h^{0,7},h^{1,6},h^{2,5},h^{3,4},h^{4,3},h^{5,2},h^{6,1},h^{7,0}]=[0,0,1,\beta,\beta,1,0,0],
\end{align} 
with $\beta = 84$, as shown in Ref. \cite{Candelas:1993nd}.  
There is a unique nontrivial element of $H^{5,2}$ which would correspond to the three-form $\Omega_{3,0}$ of an ordinary CY threefold. 
Here, the $\Omega_{5, 2}$ can be constructed by the usual way called Atiyah-Bott-G\r{a}rding-Candelas method \cite{atiyah1973lacunas, Candelas:1987se}, but at this time, it is not holomorphic. (See for details, Refs. \cite{Griffiths:1973-1, Griffiths:1973-2}.) 
In addition, $H^{4,3}$ corresponds to $H^{2,1}$ of a CY threefold, which parametrizes the complex structure deformations. 
The explicit complex structure moduli will be introduced later.

As we described, it was known that there are other possibilities to divide $1^9/\mathbb{Z}_3$ Gepner model. For instance, we can utilize the other $\mathbb{Z}_3^\prime$ action as $G$:
\begin{align}
  \mathbb{Z}_3^\prime: \left( y_1, y_2, y_3, y_4, y_5, y_6, y_7, y_8, y_9 \right) \rightarrow \left( \omega y_1, \omega y_2, \omega y_3, \omega^2 y_4, \omega^2 y_5, \omega^2 y_6, y_7, y_8, y_9 \right),
\end{align}
leading to the mirror of the rigid CY $\mathbb{P}_8[3]/\mathbb{Z}_3^\prime$. 
In this case, there exist totally $h^{4,3}=\beta = 36$ deformations, but the degree of freedom of complex structure corresponds to $30$ deformations of the equation $W=0$, maintaining the homogeneity of $W$ and the invariance under the $\mathbb{Z}_3^\prime$ symmetry. 
Similarly, we can consider the case that the $1^9/\mathbb{Z}_3$ model is further divided by an additional $\mathbb{Z}_3^{\prime\prime}$ symmetry, i.e., $\mathbb{P}_8[3]/(\mathbb{Z}_3^\prime\times \mathbb{Z}_3^{\prime\prime})$ with $G = \mathbb{Z}_3^\prime\times \mathbb{Z}_3^{\prime\prime} $ whose Hodge number is $h^{4,3}=\beta = 12$.

A geometrical interpretation of non-geometric internal CFT has also been studied. 
The structure of toroidal orbifold appears in the mirror $\tilde{\mathscr{Z}}$, and it can be understood by LG-CY correspondence from the $1^9/\mathbb{Z}_3$ Gepner model. 
Indeed, integrating out three of nine fields in Eq. (\ref{eq:1^9 superpotential}) restricts us to reside in $\mathbb{CP}^2 \times \mathbb{CP}^2 \times \mathbb{CP}^2$. This is equivalent to the product of three tori $T^2$ with the $SU(3)$ root lattice \cite{Giveon:1990ay}, although the orbifold group is non-geometrically acting symmetry \cite{Vafa:1989xc, Becker:2006ks}. 
This is a consequence of the fact that the original rigid CY $\mathscr{Z}$ is geometrically represented by a product of three tori $T^2$ subject to $\mathbb{Z}_3\times \mathbb{Z}_3^\prime$ identification \cite{Giveon:1990ay}. 
Hence, it is natural to expect that the effective action can also be accessed by T-duality from the geometric IIA model on the rigid CY, $T^6/{(\mathbb{Z}_3 \times \mathbb{Z}_3^\prime)}$ \cite{DeWolfe:2005uu}. 
Remarkably, for the $\mathbb{P}_8[3]$ case, the (2,2)-CFT description of the flux compactification was developed in Ref. \cite{Becker:2006ks}, in which they introduced O-planes with appropriate orientifold actions besides three-form fluxes and D-branes. 
These are necessary ingredients to reduce $\mathcal{N}=2$ SUSY by half and cancel a positive contribution to the D3-brane charge induced by SUSY-preserving three-form fluxes and the existence of D3-branes, i.e., the tadpole cancellation condition. 
The authors of Ref. \cite{Becker:2006ks} explicitly demonstrated several orientifold actions, and it turned out that the maximal value of the tadpole charge is 12. 

In this paper, we will focus on the mirror of rigid CY $\tilde{\mathscr{Z}}=\mathbb{P}_8[3]/\mathbb{Z}_3$ treated in Ref. \cite{Becker:2006ks}, and we adopt the orientifold action leading to the tadpole charge 12. It is notable that one can do the same calculation for the different orientifold actions since they are defined on $\mathbb{P}_8[3]$. As discussed in the next section, the $\tilde{\mathscr{Z}}$ equips a symplectic structure which allows us to employ the technique of flux compactifications on CY manifolds. There exist two types of closed string moduli, i.e., the untwisted (bulk) moduli and the twisted moduli from the viewpoint of 
toroidal orbifolds. 
The stabilization of the untwisted moduli at the AdS minima was discussed in the context of flux compactification \cite{Becker:2006ks,Becker:2007dn}, but the classification of flux vacua has not been developed yet. 
Our purpose is to explore the vacuum structure of untwisted moduli and inspect several swampland conjectures. 
The effective action of untwisted moduli is derived by calculating the period vector in Ref. \cite{Candelas:1993nd}. 
Finally, we emphasize that this background is not in one-to-one correspondence with the T-dual of DGKT model due to the different orbifolding and orientifold actions, but it captures a part of the vacuum structure discussed in Type IIA side\footnote{As long as we consider only bulk modes, these two systems are exactly dual.}.

\subsection{Type IIB flux compactifications with special geometry}
\label{sec:flux compactification with special geometry}

In this section, we briefly review Type IIB flux compactifications with special geometry. On manifolds with special geometry, a unique holomorphic three-form $\Omega$ plays an important role in introducing three-form fluxes in Type IIB string theory. 
As we mentioned in the previous section, we can identify the holomorphic three-form $\Omega$ of CY threefold with the $(5,2)$-form on $\tilde{\mathscr{Z}}$. 
In particular, we deal with untwisted cycles on 3-tori whose period integrals were explicitly calculated in Ref. \cite{Candelas:1993nd}. 

On CY threefolds $\mathscr{M}$, we can set the three-form basis  $(A^a, B_b)$ of $H_3(\mathscr{M}, \mathbb{C})$ to be symplectic because of the existence of special geometry\footnote{See for the details about the special geometry, Refs. \cite{Strominger:1990pd,Candelas:1990pi} and the relation to the CP symmetry, Ref. \cite{Ishiguro:2020nuf}.}, i.e., it satisfies the following relations:
\begin{alignat}{4}
  A^a \cap B_b &= \delta^a_b,~& B_b \cap A^a &= -\delta_b^a,~&
  A^a \cap A^b &= 0,~& B_a \cap B_b &= 0,
\end{alignat}
with $a, b = 0, 1, \dots, h^{2,1}$.
The dual cohomology basis $(\alpha_a, \beta^b)$ of $H^3(\mathscr{M}, \mathbb{C})$ is defined to satisfy
\begin{alignat}{2}
  \int_{A^a} \alpha_b = \int_{\mathscr{M}} {\alpha_b \wedge \beta^a} = {\delta^{a}}_{b},~& \int_{B_a} \beta^{b} = \int_{\mathscr{M}} {\beta^{b} \wedge \alpha_{a}}= - {\delta^{b}}_{a}, \label{eq:dual cohomology basis}
\end{alignat}
from which $(\alpha_{a}, \beta^{b})$ are the Poincar\'{e} dual with each other.
Then, we introduce the $(2h^{2,1}+2, 1)$ period vector as
\begin{align}
  \Pi \equiv
  \begin{pmatrix}
    F_a\\
    u^a
  \end{pmatrix}
  \equiv
  \begin{pmatrix}
    \int_{B_a} \Omega\\
    \int_{A^a} \Omega
  \end{pmatrix},
\end{align}
where $\{u^a\}$ denotes the projective coordinates of the complex structure moduli space and each $F_a$ becomes a function of $\{u^a\}$. 
This yields the following expansion of $\Omega$:
\begin{align}
  \Omega = u^a \alpha_a - F_a \beta^a.
\end{align}
Note that $\Omega$ is defined up to a holomorphic multiplication (see for details about the notation, Appendix \ref{sec:Special Geometry}). 
Since $\{u^a\}$ is denoted as the projective coordinates with the number $h^{2,1} + 1$, 
all of them do not correspond to the complex structure deformations. 
In a usual way, the flat coordinates are chosen as
\begin{align}
  z^a &= \frac{u^a}{u^0}, \quad (a \neq 0),
\end{align}
on the $u^0 \neq 0$ patch. 
Hereafter, we simply set $u^0 = 1$ so that $\{z^a\} ~(a = 1, 2, \dots, h^{2,1})$ becomes a set of dynamical fields, namely the complex structure moduli. 
The Kodaira theory \cite{kodaira2006complex, Candelas:1990pi} states that if we variate $\Omega$ with the complex structure $z^a$, it results in 
\begin{align}
  \frac{\partial \Omega}{\partial z^a} \in H^{3,0}(\mathscr{M}, \mathbb{C}) \oplus H^{2,1}(\mathscr{M}, \mathbb{C}).
\end{align}
Hence, we can define
\begin{align}
  \frac{\partial \Omega}{\partial z^a} = k_a \Omega + \chi_a, \label{eq:derivative of Omega by complex-structure}
\end{align}
with $\chi_a \in H^{2,1}(\mathscr{M}, \mathbb{C})$ being a harmonic $(2, 1)$-form, and the uniqueness of harmonic $(3, 0)$-form is employed here. 
This formula plays an essential role in this framework (see Appendix \ref{sec:Special Geometry} in more detail).

Let us discuss Type IIB flux compactifications, where the three-form fluxes $F_3$ and $H_3$ are quantized on each cycle as\footnote{Here and in what follows, we set the string length $l_s = 1$ for simplicity.}
\begin{alignat}{2}
  f^a \equiv \int_{A^a} F_3, ~& f_a \equiv \int_{B_a} F_3,\\
  h^a \equiv \int_{A^a} H_3, ~& h_a \equiv \int_{B_a} H_3,
\end{alignat}
so that $\{f^a, f_a, h^a, h_a\}$ becomes a set of integers. Here, $a$ runs from $0$ to $h^{2,1}$ again. 
From these expressions, $F_3$ and $H_3$ are expanded as
\begin{align}
  F_3 &= f^a \alpha_a - f_a \beta^a, \\ 
  H_3 &= h^a \alpha_a - h_a \beta^a.
\end{align}
When these quantized three-form fluxes are turned on certain three-cycles, they induce  the GVW superpotential 
in the 4D effective action \cite{Gukov:1999ya}
\begin{align}
  W = \int_{\mathscr{M}} G_3 \wedge \Omega,
\end{align}
with $G_3 \equiv F_3 - S H_3$. 
Here and in what follows, the reduced Planck mass is set to be unity, and we denote $S$ as the axio-dilaton whose imaginary part (${\rm Im}S$) determines the inverse of string coupling, i.e., ${\rm Im} S = g_S^{-1}$. 

We recall that the K\"ahler potential on the geometric CY threefold is given by
\begin{align}
  K = K_{\rm ad} + K_{\rm cs} + K_{\rm vol} = - k_S \log\left(-i \left(S-\bar{S}\right)\right) - \log\left(-i \int_{\mathscr{M}} \Omega \wedge \bar{\Omega} \right) - 2 \log \mathcal{V},
\end{align}
with $k_S = 1$ and $\mathcal{V}$ being the volume of CY.
However, as discussed in Ref. \cite{Becker:2007dn}, all the K\"ahler moduli for the non-geometric case in our interest are projected to the invariant point under the orbifolding on $\tilde{\mathscr{Z}}$. 
The contributions of fixed K\"ahler moduli change the axio-dilaton K\"ahler potential to
\begin{align}
  K_{\tilde{\mathscr{Z}}} = -4 \log\left(-i \left(S-\bar{S}\right)\right) - \log\left(-i \int_{\tilde{\mathscr{Z}}} \Omega \wedge \bar{\Omega} \right), \label{eq:generalK_nongeometric}
\end{align}
that is, $k_S = 4$.  
As a result, the 4D scalar potential $V$ is expressed by $K$ and $W$ as
\begin{align}
  V = e^K (K^{I\bar{J}}D_I W D_{\bar{J}}\bar{W} - 3 |W|^2),
\end{align}
with $I$ being the index of all the moduli, $K_{I\bar{J}} \equiv \partial_I \bar{\partial}_{\bar{J}} K$ being the K\"ahler metric, and $D_I W \equiv K_I W+ \partial_I W$ being the covariant derivative of superpotential $W$ and $K_I \equiv \partial_I K$. $\partial_I f$ means a derivative of $f$ with a variable which the index $I$ denotes. 
Note that the scalar potential $V$ does not become no-scale type, since there is no degree of freedom about the K\"ahler 
structure deformations.
All the above relevant quantities which we need in the actual moduli stabilization can be expressed by the period vector $\Pi$ (or the prepotential $F$) instead of $\Omega$, which is identified with the period vector of $\Omega_{5, 2}$ on $\tilde{\mathscr{Z}}$. 

We comment on what happens when the coefficient of the axio-dilaton part $k_S=4$ is different from the geometric one ($k_S=1$). Because of the difference, the $SL(2, \mathbb{Z})$ symmetry of the axio-dilaton
\begin{align}
  S \rightarrow \frac{a S + b}{c S + d}, \quad 
  \begin{pmatrix}
    F_3 \\ H_3
  \end{pmatrix}
  \rightarrow
  \begin{pmatrix}
    a & b \\
    c & d
  \end{pmatrix}
   \begin{pmatrix}
    F_3 \\ H_3
  \end{pmatrix}
\end{align} 
does not exist entirely in the effective action, since $e^K$ and $W$ transform as
\begin{align}
  e^K \rightarrow e^K |cS+d|^{2k_S}, \quad |W|^2 \rightarrow |cS+d|^{-2}.
\end{align}
Hence, only the $T$-transformation survives.

\subsection{Effective action}
\label{sec:effaction}
Here and in what follows, we examine a specific period vector calculated in Ref. \cite{Candelas:1993nd}. 
In particular, we deal with three moduli $\tau_i~ (i = 1, 2, 3)$, which reflects the untwisted complex structure moduli of the underlying toroidal orbifold. 
The period vector $\Pi$ takes the form
\begin{align}
    \Pi = 
      \begin{pmatrix}
    F_a\\
    u^a
  \end{pmatrix}
  =
    \begin{pmatrix}
         - \tau_1 \tau_2 \tau_3 
         \\ \tau_2 \tau_3   
         \\ \tau_1 \tau_3 
         \\ \tau_1 \tau_2
         \\ 1
         \\ \tau_1  
         \\ \tau_2 
         \\ \tau_3 
    \end{pmatrix}.
\end{align}
Since the period vector determines the K\"ahler potential of the complex structure moduli 
and the flux-induced superpotential
\begin{align}
\begin{split}
    K_{\rm cs} &= - \ln \left(-i \int_{\tilde{\mathscr{Z}}} \Omega \wedge \bar{\Omega} \right) 
    = -\ln \left(-i \Pi^\dagger \cdot \Sigma \cdot \Pi \right),
    \nonumber\\
    W &= \int_{\tilde{\mathscr{Z}}} G_3 \wedge \Omega =\sum_a \biggl[-(f^a-Sh^a)F_a + (f_a-Sh_a)u^a\biggl],
\end{split}
\end{align}
with 
$\Sigma = 
\begin{pmatrix}
0 & {\bf 1}_4\\
-{\bf 1}_4 & 0\\
\end{pmatrix}$
, their explicit forms are given by 
\begin{align}
  K &= -4\ln(-i (S - \bar{S})) -\ln(i (\tau_1 - \bar{\tau}_1) (\tau_2 - \bar{\tau}_2) (\tau_3 - \bar{\tau}_3) ), \label{eq:Kahlerpotential_anisotropic_orbifold} \\
  W &= (f^0 - S h^0)\tau_1 \tau_2 \tau_3 
  - \biggl[ (f^1 - S h^1) \tau_2 \tau_3 + (f^2 - S h^2) \tau_3 \tau_1 + (f^3 - S h^3) \tau_1 \tau_2 \biggl]
  \nonumber\\ 
  &+ \biggl[ (f_1 - S h_1) \tau_1 + (f_2 - S h_2) \tau_2 + (f_3 - S h_3) \tau_3  \biggl] + (f_0 - S h_0), 
  \label{eq:superpotential_anisotropic_orbifold}
\end{align}
where $\{f^{1,2,3}, f_{1,2,3}, h^{1,2,3}, f_{1,2,3}\}$ denotes the RR- and NSNS-flux quanta. 
The obtained scalar potential $V$ is almost the same as that in Type IIB flux compactifications on toroidal orbifolds, 
but with the slightly modified coefficient of $K_{\rm ad}$ with $k_S = 4$. 
When considering the isotropic toroidal background, all $\tau_i$ are identified with one modulus $\tau$. 
By rewriting $\{f^{1,2,3}, f_{1,2,3}, h^{1,2,3}, f_{1,2,3}\}$ by $\{f^1, f_1, h^1, h_1\}$, the whole flux sets are given by $\{f^0, f^1, f_0, f_1\}$ and $\{h^0, h^1, h_0, h_1\}$ for RR- and NSNS-fluxes, respectively.
The effective action in our interest reduces to be
\begin{align}
  K &=  -4 \ln \left(-i (S - \bar{S})\right) - 3\ln(-i (\tau - \bar{\tau})), \label{eq:Kahlerpotential_isotropic_orbifold}\\
  W &= (f^0 - S h^0) \tau^3 - 3 (f^1 - S h^1)\tau^2 + 3 (f_1 - S h_1) \tau + (f_0 - S h_0). 
  \label{eq:superpotential_isotropic_orbifold}
\end{align}
Moduli stabilization in this class of flux compactification was studied already in Ref. \cite{Becker:2007dn}, but our purpose 
is to clarify the flux landscape from a statistical viewpoint and explore the boundaries between the string landscape and the swampland.

Before concluding this section, we discuss the cancellation condition of the D3-brane charge. 
Since the three-form fluxes induce the D3-brane charge (tadpole charge)
\begin{align}
  N_{\rm flux} = \int_{\mathscr{M}} H_3 \wedge F_3 = - h^a f_a + f^a h_a,
\end{align}
it should be canceled by the so-called tadpole cancellation condition
\begin{align}
  N_{\rm flux} + N_{\rm D3} - \frac{1}{2} N_{\rm O3} = 0.
\end{align}
Here $N_{\rm D3}$ and $N_{\rm O3}$ denotes the number of ${\rm D3}$-branes and ${\rm O3}$-planes, respectively. 
The exact value of $N_{\rm O3}$ is determined by the orientifold action as constructed in Ref. \cite{Becker:2006ks}. 
In the following analysis, we fix the orientifold action as $\sigma_1$ in Ref. \cite{Becker:2006ks} unless we specify it. 
This restricts the maximum $N_{\rm flux}$ to 12.\footnote{Different orientifold actions give different values of maximum $N_{\rm flux}$ allowed. Since the value $12$ for $\sigma_1$ is determined in a discussion involving twisted cycles\cite{Becker:2006ks}, we should say that the orbifold limit considered here is just the case where the twisted moduli are set to be 0 everywhere. The stabilization of the twisted moduli is left for future work.}
Hence, the $N_{\rm flux}$ is upper bounded as 
\begin{align}
  N_{\rm flux} \leq 12.
\end{align}
Note that the tadpole charge $N_{\rm flux}$ is required to be positive in the supersymmetric compactifications on the geometrical case, but it does not hold in our non-geometric case even in the supersymmetric fluxes \cite{Becker:2007dn}.

\section{Vacuum Structure and Role of Tadpole Charge \texorpdfstring{$N_{\rm flux}$}{Nflux}}
\label{sec:vacua structure}

In this section, we analytically examine the characteristic features of the flux vacua to find out the boundaries between the landscape and the swampland. 
From a general structure of the scalar potential shown in section \ref{sec:signV}, 
we find a strong correlation between the sign of the cosmological constant $\Lambda$ and the tadpole charge $N_{\rm flux}$ due to the fact 
that the tadpole charge $N_{\rm flux}$ emerges in the scalar potential as the coefficient of the axio-dilaton. 
Furthermore, the relation between the $N_{\rm flux}$ 
and the typical moduli mass is exemplified in a concrete example in section \ref{sec:Nfluxmoduli}.

\subsection{Sign of the cosmological constant \texorpdfstring{$\Lambda$}{\unichar{"039B}} and \texorpdfstring{$N_{\rm flux}$}{Nflux}}
\label{sec:signV}
In this section, we focus on the sign of cosmological constant $\Lambda$ against the tadpole charge $N_{\rm flux}$.  
Let us define $V \equiv e^K \tilde{V}$ whose expression is useful to analytically study the vacua structure. 
It is remarkable that $V$ and $\partial_X \tilde{V}$ at the minima $\partial_X V=0$ are related as
\begin{align}
  V = - e^K \frac{\partial_{X} \tilde{V}}{K_{X}},
\end{align}
where $X$ denotes arbitrary moduli fields. 
Since $K_S < 0$ holds for the axio-dilaton direction throughout the discussion below, 
the sign of the cosmological constant is determined by $\partial_{X} \tilde{V}$ at the minima, and then it is important to consider the explicit form of $\partial_{X} \tilde{V}$. 
Recall that $V$ is always a quadratic function of $S$ regardless of the internal background $\mathscr{M}$ since $G_3 = F_3 - S H_3$ is a linear function for $S$. 
For that reason, we concentrate on derivatives of $V$ with respect to $S$ to extract some necessary conditions of the perturbatively stable vacua in a subsequent discussion.

The second derivative of $V$ with respect to ${\rm Im} S$ at the minima ($\partial_{{\rm Im}S}V=0$) is given by
\begin{align}
  e^{-K} \partial^2_{{\rm Im}S} V = \frac{1-k_S}{{\rm Im}S} \partial_{{\rm Im}S} \tilde{V} + \partial^2_{{\rm Im}S} \tilde{V},
\end{align}
where $k_S$ denotes the coefficient of the axio-dilaton K\"ahler potential $K_{\rm ad}=-k_s\ln (-i(S-\bar{S}))$. 
As mentioned before, $k_S = 1$ and 4 correspond to 
geometric CY threefolds and the non-geometric background $\tilde{\mathscr{Z}}$, respectively. 
Since
\begin{align}
  \partial_{{\rm Im}S} \tilde{V}= \frac{k_S}{{\rm Im}S} e^{-K} V
\end{align}
holds at the minima, we arrive at the necessary condition to be stable $\partial^2_{{\rm Im}S} V > 0$,\footnote{This follows from Sylvester's criterion. Since the mass matrix is real and symmetric in this case, $\partial^2_{{\rm Im}S} V$ has to be positive definite to be stable at Minkowski and dS vacua. The AdS vacua are analyzed in the next section.}
\begin{alignat}{2}
  V &< \frac{\left({\rm Im}S\right)^2}{k_S ( k_S - 1 )} e^K \partial^2_{{\rm Im}S}\tilde{V} ~ & (k_S &\neq 1), \label{eq:relation_sdVandV}\\
  0 &< \partial^2_{{\rm Im}S} \tilde{V} ~ & (k_S &= 1).
\end{alignat}
We observe that in  the $k_S \neq 1$ case of our interest, the signs of $V$ and $\partial^2_{{\rm Im}S} \tilde{V}$ at the minima are correlated with each other. If there exists a stable dS minimum ($V>0$), Eq. (\ref{eq:relation_sdVandV}) indicates that $\partial^2_{{\rm Im}S}\tilde{V}$ must be positive there. On the other hand, in the $k_S=1$ case, $\partial^2_{{\rm Im}S}V$ and $\partial^2_{{\rm Im}S} \tilde{V}$ have the same sign. In both cases, the explicit form of $\partial^2_{{\rm Im}S} \tilde{V}$ is an important ingredient in whether the minima are stable.

These fundamental observations motivate us to study the structure of $\tilde{V}$, but it is quite difficult to analyze it without any assumptions. Hence, we assume that there is no mixing between the axio-dilaton $S$ and complex structure moduli in the K\"ahler metric, which is broadly applicable to the effective action in the regime of small string coupling. 
Hereafter, we concentrate on the $k_S = 4$ case whose K\"ahler potential is given in Eq. (\ref{eq:generalK_nongeometric}), but the following results hold for the $k_S = 1$ case similarly. 

Let us split $\tilde{V}$ into the following three pieces under the above assumption;
\begin{align}
  \tilde{V}_S &= K^{S\bar{S}} D_S W D_{\bar{S}} \bar{W},\\ 
  \tilde{V}_{\rm cs} &= \sum_{a, b} K^{a\bar{b}}D_a W D_{\bar{b}} \bar{W},\\
  \tilde{V}_{\rm SG} &= - 3 |W|^2,
\end{align}
where $a, b$ run over the complex structure moduli. 
Then, we derive the quantity $\partial_{{\rm Im}S}\tilde{V}$ whose sign is directly related to the sign of the cosmological constant. (For the detailed calculation, see Appendix \ref{app:cal}.) 
It turned out that $\tilde{V}$ is of the form
\begin{align}
  \tilde{V} = \frac{1}{2} \partial^2_{{\rm Im}S} \tilde{V} ({\rm Im}S)^2 - e^{-K_{\rm cs}} N_{\rm flux} {\rm Im} S + C, \label{eq:The scalar potential in terms of Nflux}
\end{align}
for both the $k_S = 4$ and the no-scale type $k_S = 1$. Here, $C$ denotes the positive-definite function appearing in the zeroth-order part of $\tilde{V}$ with respect to ${\rm Im}S$ as in Eq. (\ref{eq:C}), and $\partial^2_{{\rm Im}S} \tilde{V}$ is explicitly written in Eq. (\ref{eq:derivative of tildeV with ImS}). 
The appearance of $N_{\rm flux}$ in the scalar potential as the coefficient of ${\rm Im}S = g_s^{-1}$ is a common feature in the no-scale type $k_S = 1$ and the non-geometric case $k_S=4$, but it is a non-trivial check to derive $N_{\rm flux}$ in $k_S=4$ case from the 10D supergravity action on non-geometric background. This expression is important to understand the 10D consistency condition on the non-geometric background, but we leave it for future work.

Following these results, we find the strong correlation between the sign of $V$ at the minimum and $N_{\rm flux}$\footnote{Although this work was carried out independently, the fact that the $N_{\rm flux}$ emerges in the scalar potential in the type IIB non-geometric flux compactification on CY manifolds was pointed out in Ref. \cite{Plauschinn:2020ram}. However, we emphasize that this emergence on the mirror of the rigid CY is still non-trivial. Furthermore, we will explore the explicit landscape on this background in analytic and numerical ways to show that there are some implications on physical quantities and the swampland conjectures from the structure of Eq. (\ref{eq:The scalar potential in terms of Nflux}).}. 
If there exists a Minkowski minimum ($V=0$), the VEV of ${\rm Im}S$ is given by
\begin{align}
  \left\langle {\rm Im}S \right\rangle = + \frac{N_{\rm flux}}{\partial^2_{{\rm Im}S} \tilde{V}}.
  \label{eq:ImS for Minkowski minima}
\end{align}
Since $\partial^2_{{\rm Im}S}V > 0$ is in one-to-one correspondence with $\partial^2_{{\rm Im}S}\tilde{V} > 0$ at $V=0$, $N_{\rm flux} > 0$ is required to obtain stable Minkowski minima. 
In general, the expectation value of ${\rm Im}S$ at an extremum of $V$ is found as
\begin{align}
  \left\langle {\rm Im}S \right\rangle = e^{-K_{\rm cs}} \frac{3N_{\rm flux} \pm \sqrt{(3N_{\rm flux})^2 - 16 C e^{2K_{\rm cs}} \partial^2_{{\rm Im}S} \tilde{V}}}{2 \partial^2_{{\rm Im}S}\tilde{V}},
  \label{eq:ImSvevgeneral}
\end{align}
from which $N_{\rm flux}$ must satisfy
\begin{align}
  (e^{-K_{\rm cs}}N_{\rm flux})^2 > \frac{16}{9} C \partial^2_{{\rm Im}S} \tilde{V}.
  \label{eq:Nfluxlower}
\end{align}
The stability condition (\ref{eq:relation_sdVandV}) states that $\partial^2_{{\rm Im}S} \tilde{V} > 0$ is required to obtain  stable Minkowski or dS minima. 
After some calculation, we find that the negative sign in Eq. (\ref{eq:ImSvevgeneral}) is a solution for the dS minima, and then $N_{\rm flux} > 0$ is also required at the stable dS minima in a similar to the Minkowski minima.
Together with Eqs. (\ref{eq:Nfluxlower}) and (\ref{eq:relation_sdVandV}), the allowed range of $N_{\rm flux}$ is further constrained to reside in
\begin{align}
  \frac{16}{9} C \partial^2_{{\rm Im}S} \tilde{V} \leq  (e^{-K_{\rm cs}} N_{\rm flux})^2 < 2 C \partial^2_{{\rm Im}S} \tilde{V}.
\end{align}

As pointed out before, it was well known that $N_{\rm flux}$ is bounded from above since only O-planes determined by the orientifold action have negative contributions to the tadpole cancellation condition, 
but our analyses show that 
the tadpole charge $N_{\rm flux}$ is further constrained by the stability condition to allow the existence of Minkowski and dS vacua. 
Hence, it suggests that it would be difficult to realize the stable dS vacua inside the string landscape due to the 
small range of $N_{\rm flux}$, in particular, the maximum number is $12$ in the orientifold action of our interest. 
On the other hand, a large $N_{\rm flux}$ contribution would allow the stable dS minima, although it is outside the landscape, 
i.e., the swampland. 
These observations will be confirmed in our numerical results in section \ref{sec:test}, in which we find the 
stable dS vacua in a region $N_{\rm flux} > 12$, indicating the strong correlation between the existence of dS vacua and the upper bound of $N_{\rm flux}$.

\subsection{\texorpdfstring{$N_{\rm flux}$}{Nflux} and moduli mass on SUSY AdS/Minkowski vacua}
\label{sec:Nfluxmoduli}

In this section, we analytically examine the structure of SUSY AdS/Minkowski vacua, paying attention to 
the relation between the $N_{\rm flux}$ and the moduli mass. 
These analyses are useful to understand the hidden structure of the swampland conjectures as analyzed in section \ref{sec:test}. (For the detail of SUSY AdS vacua, see, Ref. \cite{Becker:2007dn}.) 

We begin with the SUSY vacua of the scalar potential $V$ without no-scale structure. 
The hermitian mass matrix at $D_IW=0$ for arbitrary fields $\phi^I$ takes rather a simple form \cite{Becker:2007dn};
\begin{alignat}{2}
  \partial_{\bar{J}} \partial_{I} V &= e^K \left( D_I D_K W D_{\bar{J}} D_{\bar{L}} \bar{W} K^{K \bar{L}} - 2 K_{I \bar{J}} |W|^2 \right), \quad &
  \partial_{J} \partial_{I} V &= - e^K (D_J D_I W) \bar{W}, \label{eq:elements of mass matrix}
\end{alignat}
where $I, J, K, L$ run over the axio-dilaton and all the complex structure moduli of our interest. 
Since the following expressions hold at SUSY vacua:
\begin{alignat}{2}
  D_S D_S W &= \frac{k_S (1 - k_S)}{(S - \bar{S})^2} W, \quad & D_S D_a W &= - D_a W_{\rm NS},  \nonumber \\
  D_a D_b W &= - i e^{K_{\rm cs}} {\kappa_{a b}}^{\bar{c}} (\bar{S} - S) D_{\bar{c}}\overline{W}_{\rm NS},&
\end{alignat}
they enable us to simplify the mass matrix there. The value of $k_S$ affects only $D_S D_S W$, and it is identically zero in the no-scale scalar potential ($k_S = 1$).

To find the analytical expression of the moduli mass, we focus on the isotropic background $\tau=\tau^1=\tau^2=\tau^3$. 
At the SUSY AdS vacua, the physical mass matrix becomes
\begin{align}
  \frac{ M^2_{\rm phys, AdS} }{\Lambda_{\rm AdS}} &= 
  \begin{pmatrix}
      \frac{2}{3} - \frac{19}{108} |x|^2 & \frac{2}{9} \bar{y} & - \frac{x}{2\sqrt{3}} - \frac{x \bar{y}}{9 \sqrt{3}} & \frac{\bar{x}}{6 \sqrt{3}} \\
      \frac{2}{9} y & \frac{2}{3} - \frac{19}{108} |x|^2 & \frac{x}{6\sqrt{3}} & - \frac{x}{2 \sqrt{3}} - \frac{\bar{x} y}{9 \sqrt{3}} \\
      - \frac{x}{2\sqrt{3}} - \frac{\bar{x}y}{9\sqrt{3}} & \frac{\bar{x}}{6 \sqrt{3}} & - \frac{7}{3} - \frac{1}{36} |x|^2 & 1 \\
      \frac{x}{6 \sqrt{3}} & - \frac{\bar{x}}{2\sqrt{3}} - \frac{x \bar{y}}{9 \sqrt{3}} & 1 & - \frac{7}{3} - \frac{1}{36} |x|^2 
  \end{pmatrix}
  ,
\end{align}
where $\Lambda_{\rm AdS} = - 3 e^K |W|^2 = - 3 e^{K_{\rm ad} + K_{\rm cs}} |W|^2$, and $x, y$ are defined by
\begin{align}
  x &\equiv (S - \bar{S})(\tau - \bar{\tau}) \frac{D_\tau W_{\rm NS}}{W}, \\
  y &\equiv (S - \bar{S})(\tau - \bar{\tau}) \frac{\overline{D_\tau W_{\rm NS}}}{W},
\end{align}
satisfying $|x| = |y|$.
The basis of mass matrix basis is set as $ \{ \bar{\hat{\tau}}, \hat{\tau}, \bar{\hat{S}}, \hat{S} \}$ for the row and $\{ \hat{\tau}, \bar{\hat{\tau}}, \hat{S}, \bar{\hat{S}} \}$ for the column. Here, 
$\hat{S}$ and $\hat{\tau}$ are 
canonically normalized as
\begin{align}
  \hat{S} &= \frac{1}{\langle {\rm Im}S \rangle} \Delta S, \\
  \hat{\tau} &= \frac{\sqrt{3}}{2} \frac{1}{\langle {\rm Im}\tau \rangle} \Delta \tau,
\end{align}
with $S = \langle S \rangle + \Delta S$ and $\tau = \langle \tau \rangle + \Delta \tau$ (in what follows, we omit $\langle ~ \rangle$ that denotes a VEV). Hence, we can see that the magnitude of the mass matrix is controlled by the parameter $x$ as stated in Ref. \cite{Becker:2007dn}. 
To simplify our analysis, let us assume that the phases of $x$ and $y$ are 0 and take
\begin{align}
r = |x| = |y|.
\end{align}
It is found that the mass squared of lightest modulus in the real basis is given by
\begin{align}
m^2_{\rm light} =\frac{\Lambda_{\rm AdS}}{2\times 108} \biggl[12 r - 11 r^2 + 4\left(-9 + \sqrt{(3 + r)^2 (81 - 36 r + 7 r^2)}\right) \biggl],
\end{align}
whose expression is useful to understand the structure of swampland conjectures in section \ref{sec:test}.

Interestingly, $x$ and $y$ also control the magnitude of $N_{\rm flux}$ as analyzed in detail below. 
Recall that three-forms $\{ F_3, H_3\}$ and $N_{\rm flux}$ are expanded as in Eqs. (\ref{eq:expansionFH}) 
and (\ref{eq:expansionNflux}) respectively, 
the SUSY conditions enable us to simplify the expression of $N_{\rm flux}$. 
We find that the $N_{\rm flux}$ at the SUSY vacua is given in Eq. (\ref{eq:NfluxAdS}) in a general complex structure 
moduli space leading to the non-diagonal K\"ahler metric $K_{a \bar{b}}$. (For more details, see, Appendix \ref{app:Nflux}.) 
When we restrict ourselves to the isotropic complex structure modulus in a similar to the previous analysis, 
Eq. (\ref{eq:NfluxAdS}) reduces to
  \begin{align}
      \frac{N_{\rm flux}}{\Lambda_{\rm AdS}} =  \frac{8 ({\rm Im}S)^3}{3} \left( 8 - \frac{|x|^2}{3}\right).
      \label{eq:NfluxAdSiso}
  \end{align}
In this way, $N_{\rm flux}$ also depends on $|x|$, and there should be a relation between the mass squared of lightest modulus $m_{\rm light}^2$ and $N_{\rm flux}$. 

To discuss the relation, we rewrite Eq. (\ref{eq:NfluxAdSiso}) as
  \begin{align}
      |x|^2 = 24 - \frac{9N_{\rm flux}}{8({\rm Im}S)^2\Lambda_{\rm AdS}}. \label{eq:xsquared}
  \end{align}
When we assume $8({\rm Im}S)^2\Lambda_{\rm AdS}\gg 9N_{\rm flux}$ due to the small contribution from $N_{\rm flux}$, the lightest modulus mass and the AdS scale 
are related as
  \begin{align}
      m^2_{\rm light} \simeq \frac{\Lambda_{\rm AdS}}{4},
   \end{align}
namely
  \begin{align}
      |m_{\rm light}R_{\rm AdS}| \simeq  \frac{\sqrt{6}}{2}\simeq 1.22.
      \label{eq:AdS/moduli_analytic}
   \end{align}
Here, we introduce the AdS scale on the $d$-dimensional spacetime as
  \begin{align}
     R_{\rm AdS} = \sqrt{\frac{(d-1)(d-2)}{|\Lambda_{\rm AdS}|}},
   \end{align}
with $d=4$. 
It suggests that the AdS/moduli mass separation conjecture holds on the isotropic 
moduli space with the definite ${\cal O}(1)$ parameter. 
The sub-leading term in Eq. (\ref{eq:AdS/moduli_analytic}) is determined by $N_{\rm flux}$. 
As will discussed in section \ref{sec:test}, 
the numerical results support these results.

We move on to the analysis of the SUSY Minkowski vacua. 
Analogously, it enables us to analyze the SUSY Minkowski vacua at which the physical mass matrix is evaluated as
\begin{align}
  \frac{ M^2_{\rm phys, Minkowski} }{e^K} &= 
   \begin{pmatrix}
       \frac{19}{72} |\hat{x}|^2 & 0 & \frac{\hat{x}^2}{6\sqrt{3}} & 0 \\
       0 & \frac{19}{72} |\hat{x}|^2 & 0 & \frac{\bar{\hat{x}}^2}{6\sqrt{3}} \\
       \frac{\bar{\hat{x}}^2}{6 \sqrt{3}} & 0 & \frac{|\hat{x}|^2}{24} & 0 \\
       0 & \frac{\hat{x}^2}{6 \sqrt{3}} & 0 & \frac{|\hat{x}|^2}{24}
   \end{pmatrix}
   ,
   \label{eq:M2Minkowski}
\end{align}
with 
\begin{align}
  \hat{x} \equiv (S - \bar{S}) (\tau -\bar{\tau}) D_{\tau} W_{\rm NS}.
\end{align}
Here, the basis of the mass matrix is the same as the analysis of AdS vacua. 
As discussed below, this expression does not depend on the value of $k_S$; thereby it is applicable 
to the scalar potential with the no-scale structure ($k_S=1$). 
The mass squared of lightest modulus in the real basis is given by
\begin{align}
m_{\rm light}^2= \frac{1}{144}(11 - 4\sqrt{7})|\hat{x}|^2,
\label{eq:mlight_Minkowski}
\end{align}
which ensures that the SUSY Minkowski vacua are always perturbatively stable when $N_{\rm flux}>0$. 

Since $W_{\rm NS} = 0 = W_{\rm RR}$ hold at the SUSY Minkowski vacua, $N_{\rm flux}$ in Eq. (\ref{eq:NfluxAdS}) 
reduces to
\begin{align}
  N_{\rm flux} &= - i e^{K_{\rm cs}} (S - \bar{S}) K^{a \bar{b}} D_{a} W_{\rm NS} D_{\bar{b}} \bar{W}_{\rm NS},
\end{align}
and it is further simplified on the isotropic background, 
  \begin{align}
      \frac{N_{\rm flux}}{e^K} = \frac{8 ({\rm Im}S)^3}{3} \frac{|\hat{x}|^2}{3}.
  \end{align}
The mass squared of lightest modulus and $N_{\rm flux}$ are in turn related as
\begin{align}
m_{\rm light}^2= \frac{11 - 4\sqrt{7}}{128}\frac{N_{\rm flux}}{({\rm Im}S)^2}.
\end{align}

Before concluding this section, let us compare the above results to the no-scale scalar potential corresponding to the geometric toroidal background with $k_S=1$. Since the no-scale structure does not lead to the AdS vacua, we focus on the 
SUSY Minkowski vacua. 
The hermitian mass matrix in the no-scale scalar potential $V$ becomes
\begin{alignat}{2}
  \partial_{\bar{J}} \partial_{I} V &= e^K \left( D_I D_K W D_{\bar{J}} D_{\bar{L}} \bar{W} K^{K \bar{L}} + K_{I \bar{J}} |W|^2 \right), \quad &
  \partial_{J} \partial_{I} V &= + 2e^K (D_J D_I W) \bar{W}, 
  \label{eq:elements of mass matrix with no-scale V}
\end{alignat}
and in particular, it is of the same form (\ref{eq:M2Minkowski}) 
at the SUSY Minkowski minima. 
Note that the $N_{\rm flux}$ in the $k_S = 1$ case is given by
  \begin{align}
      \frac{N_{\rm flux}}{e^K} = + \frac{|\hat{x}|^2}{3}.
  \end{align}
Then, we find that the lightest modulus mass (\ref{eq:mlight_Minkowski}) is determined by $N_{\rm flux}$,
\begin{align}
m_{\rm light}^2= \frac{11 - 4\sqrt{7}}{48}N_{\rm flux}.
\end{align}
Remarkably, the mass squared of lightest modulus is proportional to $N_{\rm flux}$ in 
both the geometric ($k_S=1$) and non-geometric ($k_S=4$) cases. 
From these expressions, the SUSY Minkowski vacua are perturbatively stable 
when $N_{\rm flux}>0$ which is consistent with the analysis in section \ref{sec:signV}.

\section{Inspection of Swampland Conjectures}
\label{sec:test}

We numerically investigate the distributions of flux vacua and 
compare them with three types of swampland conjectures; the 
AdS/moduli scale separation conjecture in section \ref{sec:AdS/moduli}, 
the AdS distance conjecture in section \ref{sec:AdSdistance} 
and the dS swampland conjecture in section \ref{sec:dS}.

\subsection{Distributions of the flux vacua}
\label{sec:distributions}

We numerically searched the flux vacua in two cases; (i) the isotropic moduli whose K\"ahler potential and superpotential are given by Eqs. (\ref{eq:Kahlerpotential_isotropic_orbifold}) and (\ref{eq:superpotential_isotropic_orbifold}) 
and (ii) the anisotropic case whose K\"ahler potential and superpotential are given by Eqs. (\ref{eq:Kahlerpotential_anisotropic_orbifold}) and (\ref{eq:superpotential_anisotropic_orbifold}), 
respectively. We directly minimized the scalar potential by utilizing the ``FindRoot''  function in Mathematica (v12.0). 
Although the various toroidal orbifolds have been discussed in the literature, the value of $k_S$ is different from the usual geometric case, as we already mentioned. 
To perform the numerical search in isotropic and anisotropic cases, we randomly generated $10{,}129{,}591$ and $17{,}136{,}095$ set of fluxes within
\begin{align}
 -20 \leq  \{f^0, f^1, f_0, f_1, h^0, h^1, h_0, h_1\} \leq 20,
\end{align}
taking into account the tadpole cancellation condition
\begin{align}
  -12 \leq N_{\rm flux} \leq 12,
\end{align}
respectively\footnote{Theoretically, there is no lower bound on the value of $N_{\rm flux}$. Nevertheless, we set it to -12 because of the amount of memory required to run the numerical calculations. Also, if we increase the value of $|N_{\rm flux}|$, it will become impossible to ignore the effects of corrections, which are left for future studies. Hence, we believe that the choice is not an irrational setting. }. 
As summarized in Tables \ref{tab:Numberofvacua_isotropic_orbifold} and \ref{tab:Numberofvacua_anisotropic_orbifold}, there exist SUSY and non-SUSY AdS stable vacua. 
It turns out that the number of SUSY AdS vacua is much smaller than that of non-SUSY AdS vacua. 
The benchmark points for SUSY and non-SUSY AdS vacua are summarized in Table \ref{tab:The benchmark points for the orbifold limit}. Here, the stability of AdS vacuum is perturbatively ensured when the moduli masses satisfy the Breitenlohner-Freedman (BF) bound. (See for detail, e.g., Ref. \cite{Freedman:2012zz}.) 
Note that we cannot exclude Minkowski or dS vacua in our limited numerical search, but the results we obtained illustrate the tendency in the flux landscape. 

\begin{table}[ht]
   \begin{center}
   \scalebox{0.9}{
    \begin{tabular}{cc}\hline
      Type & Number of flux vacua (Stable/Unstable) \\ \hline
      SUSY AdS vacua & $232{,}800$/0 \\ 
      non-SUSY AdS vacua & $1{,}672{,}413$/$360{,}336$ \\
      SUSY Minkowski vacua & 0\\
      non-SUSY Minkowski vacua & 0 \\
      Unstable dS vacua & 0 \\
      \hline
    \end{tabular}
    }
  \end{center}
  \caption{The number of vacua in the isotropic case.}
  \label{tab:Numberofvacua_isotropic_orbifold}
\end{table}

\begin{table}[ht]
   \begin{center}
   \scalebox{0.9}{
    \begin{tabular}{cc}\hline
      Type & Number of flux vacua (Stable/Unstable) \\ \hline
      SUSY AdS vacua & 390/0 \\ 
      non-SUSY AdS vacua & $5{,}893$/$4{,}305$ \\
      SUSY Minkowski vacua & 0/0\\
      non-SUSY Minkowski vacua & 0/0 \\
      Unstable dS vacua & 0/0 \\
      \hline
    \end{tabular}
    }
  \end{center}
  \caption{The number of vacua in the anistropic case.}
  \label{tab:Numberofvacua_anisotropic_orbifold}
\end{table}

\begin{table}[ht]
\centering
\begin{tabular}{cccccc}\hline
Properties                         & Vacuum 1 & Vacuum 2 & Vacuum 3 & Vacuum 4  \\\hline
SUSY?                              & Yes       & Yes       & No        & No        \\
$f^0$                              & -1        & -8        & -10       & 7         \\
$f^1$                              & 9         & 1         & -10       & -9        \\
$f^2$                              & 7         & 10        & 10        & -4        \\
$f^3$                              & 1         & 10        & -9        & -8        \\
$f_0$                              & 4         & -4        & 2         & 1         \\
$f_1$                              & -1        & 8         & -3        & 0         \\
$f_2$                              & -2        & 2         & 8         & 9         \\
$f_3$                              & 3         & -4        & 3         & -6        \\
$h^0$                              & 1         & 0         & -1        & 0         \\
$h^1$                              & -3        & 4         & 2         & -2        \\
$h^2$                              & 0         & 5         & 8         & -5        \\
$h^3$                              & 0         & 4         & 0         & -4        \\
$h_0$                              & -8        & -9        & -10       & 7         \\
$h_1$                              & 3         & -9        & 8         & -8        \\
$h_2$                              & -1        & -3        & 1         & 4         \\
$h_3$                              & -9        & -1        & -3        & 3         \\
$\langle {\rm Re} S \rangle $      & -1.998    & 0.429     & 3.426     & -0.532    \\
$\langle {\rm Im} S \rangle $      & 3.568     & 5.649     & 17.954    & 8.141     \\
$\langle {\rm Re} \tau_1 \rangle $ & -1.166    & -1.797    & 3.012     & 0.168     \\
$\langle {\rm Im} \tau_1 \rangle $ & 1.461     & 2.42      & 3.053     & 2.867     \\
$\langle {\rm Re} \tau_2 \rangle $ & 3.568     & 0.435     & -6.872    & -2.386    \\
$\langle {\rm Im} \tau_2 \rangle $ & 3.225     & 3.956     & 3.557     & 4.075     \\
$\langle {\rm Re} \tau_3 \rangle $ & -0.343    & 0.0645    & 0.180     & -1.728    \\
$\langle {\rm Im} \tau_3 \rangle $ & 3.326     & 2.744     & 2.962     & 6.947     \\
$V$                                & -0.0929   & -0.1083   & -0.0054   & -0.04525  \\
$N_{\rm flux}$                     & 12        & -3        & 1         & 12        \\\hline
\end{tabular}
\caption{The benchmark points for SUSY and non-SUSY AdS vacua in the anisotropic case.}
\label{tab:The benchmark points for the orbifold limit}
\end{table}

\subsection{The AdS/moduli scale separation conjecture}
\label{sec:AdS/moduli}

We begin with the AdS/moduli scale separation conjecture proposed 
in Ref. \cite{Gautason:2018gln}, stating that the AdS size $R_{\rm AdS}$ and the 
mass of lightest modulus $m_{\rm light}$ satisfy the equality
\begin{align}
m_{\rm light} R_{\rm AdS} \leq c,
\label{eq:AdS/moduli}
\end{align}
in an AdS minimum. 
The constant $c$ is regarded as ${\cal O}(1)$ constant. 
That conjecture is supported by the Type IIB supergravity solution on 
$AdS_5\times S^5$, on which the mass of the lightest modulus 
satisfy $m_{\rm light}\sim R_{S^5}^{-1}\sim R_{AdS_5}^{-1}$. 
Note that the sizes of $AdS_5$ ($R_{AdS_5}$) and $S^5$ ($R_{S^5}$) 
are correlated by the 5-form fluxes supporting the supergravity solution. 

\begin{figure}[H]
\begin{minipage}{1.0\hsize}
 \begin{center}
     \includegraphics[page=1, width=170mm]{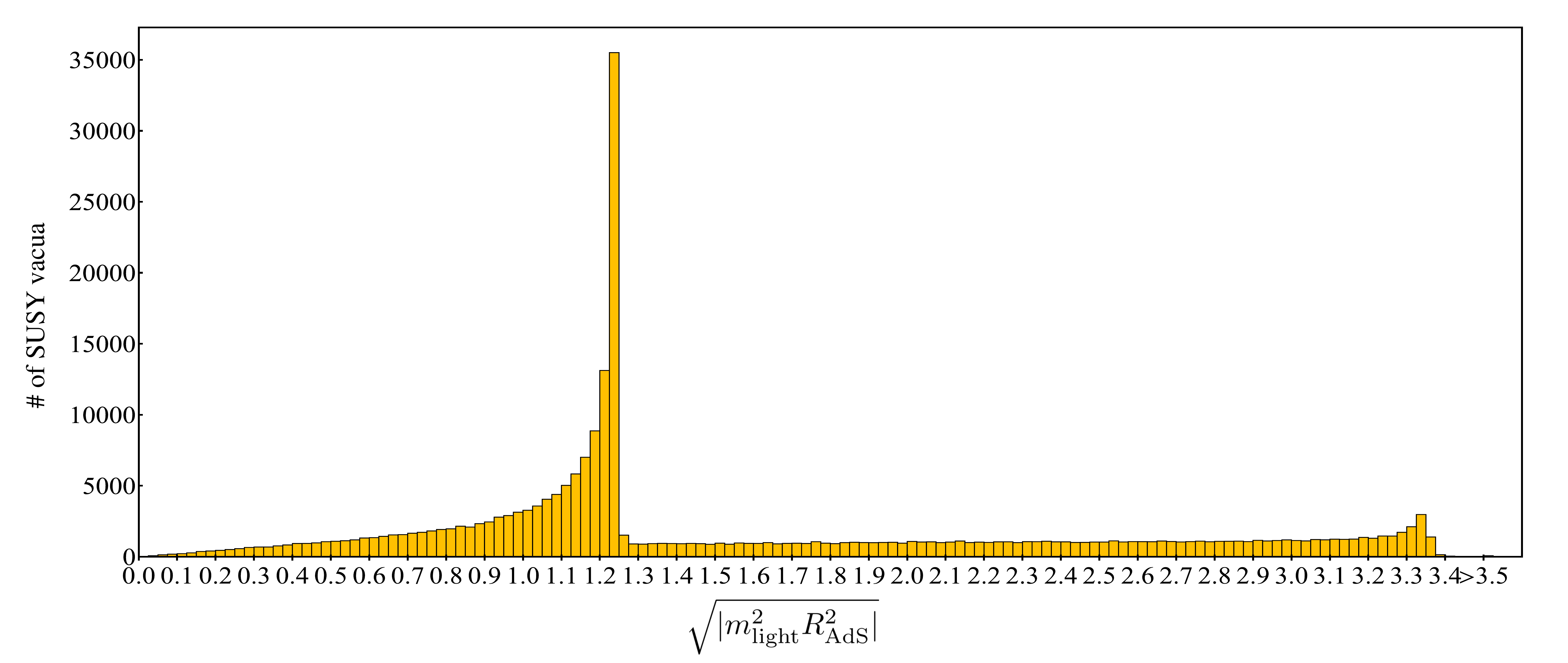}
 \end{center}
\caption{The distributions of $\sqrt{m^2_{\rm light}R^2_{\rm AdS}}$ at SUSY AdS vacua in the isotropic moduli space. All vacua here satisfy the BF bound.}
    \label{fig:scaleseparation_isotropic_SUSY}
\end{minipage}\\
\begin{minipage}{1.0\hsize}
\centering
     \includegraphics[page=2, width=170mm]{AdS_moduli_scale_separationv2_isotropic.pdf}
\end{minipage}
\caption{The distributions of $\sqrt{m^2_{\rm light}R^2_{\rm AdS}}$ at non-SUSY AdS vacua in the isotropic moduli space. All vacua satisfy the BF bound. Here, we set the increments to be 0.025 to clarify the fine structure of the distribution.}
    \label{fig:scaleseparation_isotropic_nonSUSY}
\end{figure}

\begin{figure}[H]
\begin{minipage}{1.0\hsize}
 \begin{center}
     \includegraphics[page=1,
     width=150mm]{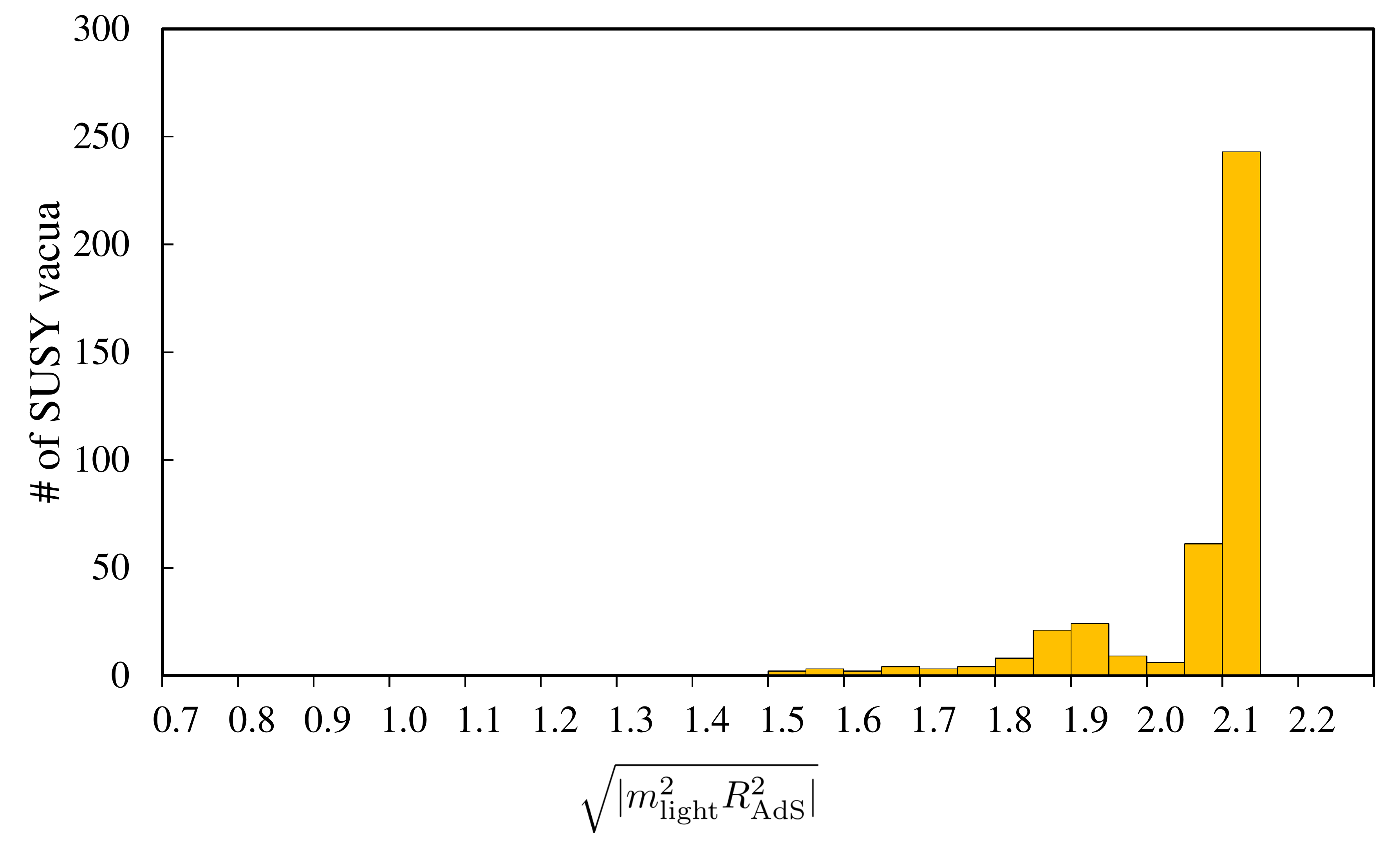}
 \end{center}
\caption{The distributions of $\sqrt{m^2_{\rm light}R^2_{\rm AdS}}$ at SUSY AdS vacua in the anisotropic moduli space. All vacua here satisfy the BF bound.}
    \label{fig:scaleseparation_anisotoropic_SUSY}
\end{minipage}
\begin{minipage}{1.0\hsize}
\centering
     \includegraphics[page=2,
     width=150mm]{AdS_moduli_scale_separationv2.pdf}
\end{minipage}
\caption{The distributions of $\sqrt{m^2_{\rm light}R^2_{\rm AdS}}$ at non-SUSY AdS vacua in the anisotropic moduli space. All vacua here satisfy the BF bound.}
    \label{fig:scaleseparation_anisotoropic_nonSUSY}
\end{figure}

In Figs. \ref{fig:scaleseparation_isotropic_SUSY} and \ref{fig:scaleseparation_isotropic_nonSUSY}, we fist plot the distributions of 
SUSY and non-SUSY AdS vacua against $\sqrt{m^2_{\rm light}R^2_{\rm AdS}}$ in the isotropic moduli space, respectively. 
Here, we employ $R_{\rm AdS} = \sqrt{\frac{(d-1)(d-2)}{|\Lambda_{\rm AdS}|}}$ with $d=4$. 
It turns out that $\sqrt{m^2_{\rm light}R^2_{\rm AdS}}$ is distributed around ${\cal O}(1)$ values, but the distribution peaks at specific values, namely ${\cal O}(1.2)$ at the SUSY vacua and ${\cal O}(1.6)$ at the non-SUSY vacua, respectively. 
The points on which the $\sqrt{m^2_{\rm light}R^2_{\rm AdS}}$ peaks are a little bit different between the SUSY and non-SUSY vacua. 
Similar phenomena appear in the distributions of the anisotropic moduli space as drawn in Figs. \ref{fig:scaleseparation_anisotoropic_SUSY} and \ref{fig:scaleseparation_anisotoropic_nonSUSY}, which correspond to the 
SUSY and non-SUSY AdS vacua, respectively. 

To clarify why the ${\cal O}(1)$ parameter $c$ in the conjecture (\ref{eq:AdS/moduli}) is sharply peaked at the specific values, we come back to the 
analytical expressions of section \ref{sec:signV}. 
On the isotropic moduli space, the product of lightest modulus mass and the AdS scale 
has a specific value $\sqrt{6}/2\simeq 1.22$ as in Eq. (\ref{eq:AdS/moduli_analytic}) under the assumption $8({\rm Im}S)^2\Lambda_{\rm AdS}\gg 9N_{\rm flux}$. 
Hence, this value is well in accord with the peak value in the numerical analysis, 
meaning that the assumption holds in this case. 
These peculiar phenomena are special to the landscape since only small $N_{\rm flux}$ is consistent with the tadpole cancellation condition. It indicates that the larger $N_{\rm flux}$ to be possible in the swampland does not lead to such a sharp distribution. 

By contrast, the distribution of non-SUSY stable AdS vacua is different from the SUSY case, although there exist several peaks in Fig. \ref{fig:scaleseparation_isotropic_nonSUSY}. 
It suggests several classes of non-SUSY AdS vacua to be consistent with the 
tadpole cancellation condition. 
For illustrative purposes, let us consider the following superpotential,
\begin{align}
W = \tau^3 +3 q\tau^2 -6q^2 \tau -S,
\label{eq:Wexamp}
\end{align} 
leading to the $N_{\rm flux}=1$ without specifying the flux quantum $q$. 
This example admits a dilute flux limit $|q|\rightarrow \infty$. 
By solving the minimum conditions $V_I=0$ with $I=S,\tau$, 
we find the non-SUSY minima
\begin{align}
{\rm Re}S={\rm Re}\tau=0,
\quad
{\rm Im}S=-16q^3,\quad
{\rm Im}\tau = -2q, \label{eq:non-SUSY solution in inspection}
\end{align} 
which requires the negative value of $q$. 
The vacuum energy is indeed negative
\begin{align}
\Lambda = \frac{1}{524288q^9}, \label{eq:non-SUSY cc in inspection}
\end{align} 
and in the large $|q|$ limit, the mass squared of the lightest modulus is evaluated as
\begin{align}
m_{\rm light}^2 \simeq -\frac{5}{54525952 q^{15}}.
\end{align} 
It turns out that the lightest modulus mass and the vacuum energy satisfy 
the relation
\begin{align}
m_{\rm light}^2  \simeq 313 |\Lambda|^{15/9}.
\end{align} 
In this way, this illustrative example satisfies the AdS/moduli scale separation conjecture in the limit $\Lambda \rightarrow 0$. 
However, the ${\cal O}(1)$ parameter $c$ in the conjecture (\ref{eq:AdS/moduli}) is not constant, but the moduli dependent. 
Such a moduli-dependent $c$ changes the structure of the distributions of non-SUSY AdS vacua in comparison with the SUSY AdS vacua as shown in Fig. \ref{fig:scaleseparation_isotropic_nonSUSY}. 
It is notable that all the non-SUSY AdS vacua in our dataset satisfy the tadpole cancellation condition, and there is no counterexample for this conjecture.

On the anisotropic moduli space, the distributions of AdS vacua are similar to the isotropic case, although they have a peek at a different value. 
We expect that this is due to the same reason as stated in the isotropic case. 
So far, the value of ${\cal O}(1)$ parameter $c$ in the conjecture (\ref{eq:AdS/moduli}) was supported by the specific Type IIB supergravity solution, but it was unclear how the ${\cal O}(1)$ parameter is distributed in the string landscape. 
By employing our method, we would reveal the hidden structure of AdS landscape. 
Indeed, our results exhibit that the distributions of  ${\cal O}(1)$ parameter have a characteristic feature in 
both the SUSY and non-SUSY vacua. 
It is interesting to find out the distribution of ${\cal O}(1)$ parameter in other corners of the string 
landscape, which we leave for future work.

\subsection{The AdS distance conjecture}
\label{sec:AdSdistance}

In this section, we deal with the AdS distance conjecture proposed in Ref. \cite{Lust:2019zwm}, 
stating that at the zero-limit of the cosmological constant $\Lambda \rightarrow 0$, 
there exists an infinite tower of light states whose masses in Planck mass units behave 
as
\begin{align}
m_{\rm tower} = c|\Lambda|^{\alpha}, \label{eq:AdS distance conjecture}
\end{align} 
where  $\alpha$ is a positive ${\cal O}(1)$ constant and $c={\cal O}(1)$. 
The conjecture was originally stated on $d$-dimensional AdS space and especially, $\alpha=1/2$ 
for supersymmetric AdS vacua in its stronger version which corresponds to 
the Maldacena-Nu$\tilde{\rm n}$ez type no-go theorem for the class of Type II AdS flux vacua without 
negative tension objects \cite{Gautason:2015tig}. 
In the following analysis, we focus on just the KK tower in our setup and discuss the consequence of that conjecture.  

To clarify our discussion, we adopt the superpotential (\ref{eq:Wexamp}) as an illustrative example again and consider the non-SUSY vacuum (\ref{eq:non-SUSY solution in inspection}). The expression of mass $m_{\rm tower}$ is taken from the T-dual picture, i.e., $m_{\rm tower}$ in Eq. (60) of Ref. \cite{Blumenhagen:2019vgj}. Taking into account the fixed value of the K\"ahler moduli on $\tilde{\mathscr{Z}}$, we find 
\begin{align}
    m^2_{\rm tower} = \frac{4}{6^{2/3}}  \frac{1}{({\rm Im}S)^2 {\rm Im \tau}}. \label{eq:lightest KK mass}
\end{align}
By substituting the minima (\ref{eq:non-SUSY solution in inspection}) into $m_{\rm tower}$ (\ref{eq:lightest KK mass}) together with Eq. (\ref{eq:non-SUSY cc in inspection}), it results in 
\begin{align}
    m^2_{\rm tower} \simeq |\Lambda|^{\frac{7}{9}},
\end{align}
thereby $\alpha = 7/18$ holds irrespective of the flux $q$. This result is consistent with the results of previous studies in the DGKT model \cite{Blumenhagen:2019vgj}. 


After the fundamental observation above, we study numerical analysis in both the isotropic and anisotropic cases. There is a subtlety in determining $\alpha$ from the numerical results that there are indeed two undefined parameters $\alpha$ and $c$ in the expression of conjecture (\ref{eq:AdS distance conjecture}). Thus, we have to fix one of them by hand, at first. 
Let us consider only SUSY AdS vacua and assume $\alpha = 1/2$ for a moment to check whether the parameter $c$ takes the $\mathcal{O}(1)$ value. 
From numerically found 232,800 and 390 SUSY AdS vacua on the isotropic and anisotropic backgrounds in Tables \ref{tab:Numberofvacua_isotropic_orbifold} and \ref{tab:Numberofvacua_anisotropic_orbifold}, 
we statistically analyzed the distributions of the parameter $c$ as summarized in Table \ref{tab:Parameter_c_alphahalf}.\footnote{We consider only the lightest KK mass associated with the three complex structure moduli.} 
It turns out that the parameter $c$ takes the $\mathcal{O}(1)$ value for all the SUSY AdS vacua with $\alpha = 1/2$, in particular, the average value and the standard deviation (SD) of $c$ are 0.134 and 0.0541 in the anisotropic case, respectively. 
However, it is still interesting to ask whether the $\alpha=1/2$ is statistically favored in the string landscape.

\begin{table}[h]
   \begin{center}
    \begin{tabular}{cccccc}\hline
       Type & Average & Median & Standard Deviation (SD) & Maximum & Minimum \\\hline
       iso.  & 0.319 & 0.258 & 0.350 & 46.3 & 0.0869  \\
       anis. & 0.134   & 0.123  & 0.0541 & 0.375   & 0.0400 \\
      \hline
    \end{tabular}
  \end{center}
  \caption{Statistical Data of $c$ for SUSY AdS vacua with $\alpha = 1/2$ for the isotropic (iso.) and anistropic (anis.) case.}
  \label{tab:Parameter_c_alphahalf}
\end{table}

\begin{figure}[H]
\begin{minipage}{1.0\hsize}
\centering
     \includegraphics[page=3, width=150mm]{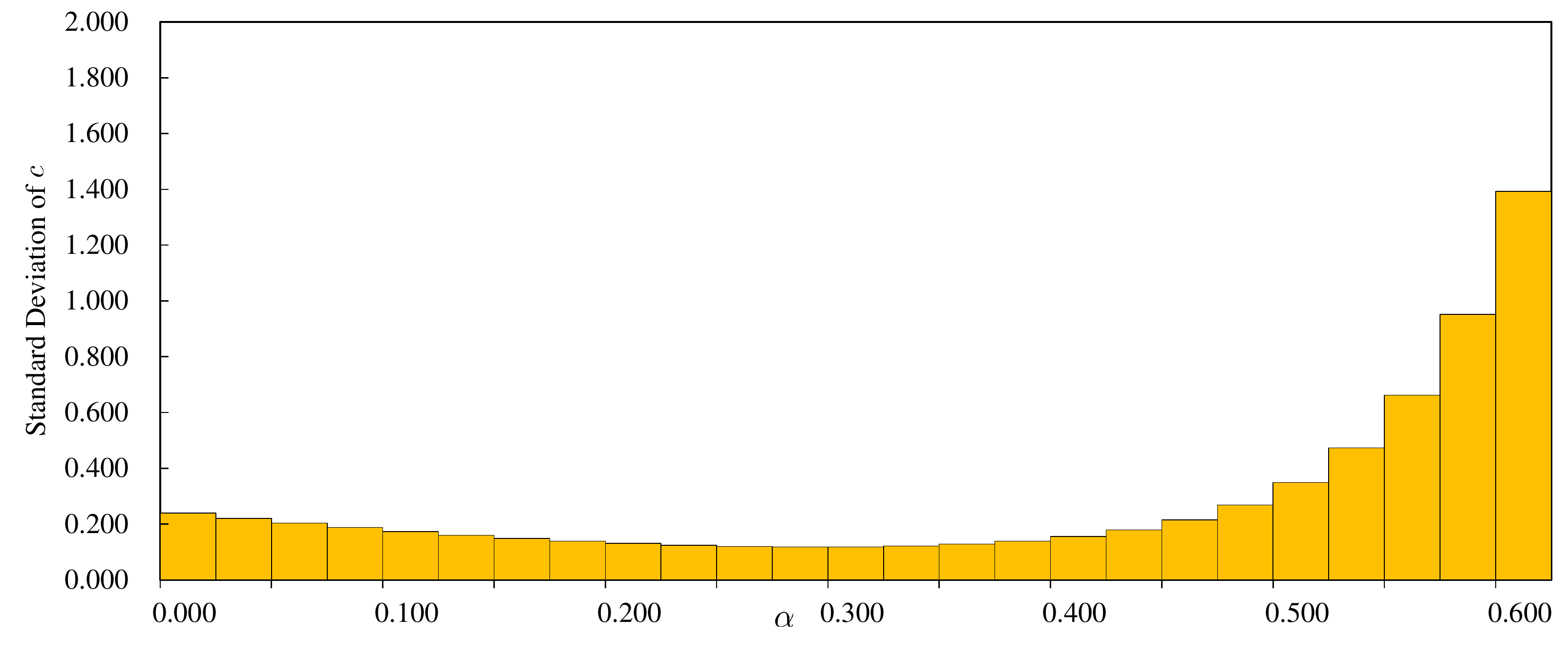}
\end{minipage}
    \caption{The standard deviation of $c = m_{{\rm tower}}/{|\Lambda|^\alpha}$ for each $\alpha$ in the case of isotropic SUSY vacua. The range of $\alpha$ shown in the figure is set as $[0.000, 0.600]$ and the increments are 0.025. In this figure, $\alpha = 0.275$ gives the minimum value $1.09 \times 10^{-2}$.}
    \label{fig:SD_isotropic_SUSY}
\end{figure}

\begin{figure}[H]
\begin{minipage}{1.0\hsize}
\centering
     \includegraphics[page=1, width=150mm]{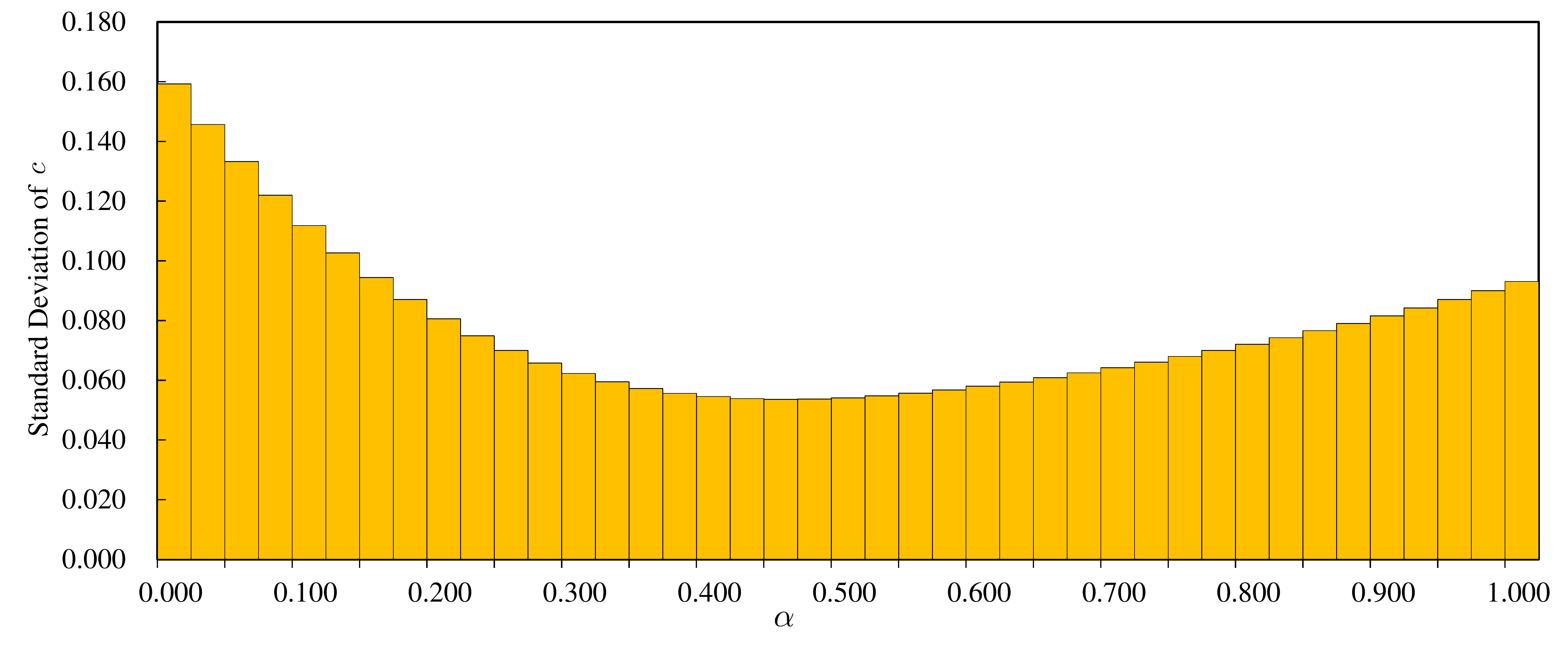}
\end{minipage}
    \caption{The standard deviation of $c = m_{{\rm tower}}/{|\Lambda|^\alpha}$ for each $\alpha$ in the case of anistropic SUSY vacua. The range of $\alpha$ shown in the figure is set as $[0.000, 1.000]$ and the increments are 0.025. In this figure, $\alpha = 0.450$ gives the minimum value $5.36 \times 10^{-2}$.}
    \label{fig:SD_anisotropic_SUSY}
\end{figure}

Thus, we change $\alpha$ and calculate SD in the same way to show the dispersion of $c$ for various $\alpha$. 
By setting the increments and the range of $\alpha$ as 0.025 and $[0.000,1.000]$, respectively, we find that $c$ is always $\mathcal{O}(1)$ value around $\alpha = 1/2$ in both the isotropic and anisotropic cases. 
Interestingly, the amount of dispersion exhibits a characteristic behavior with respect to the value of $\alpha$. From Figs. \ref{fig:SD_isotropic_SUSY} and \ref{fig:SD_anisotropic_SUSY}, it can be seen easily that the SD of $c$ varies over the curve and the minimum in the isotropic and anisotropic cases is located around $\alpha = 0.275$ and $\alpha = 0.450$, at which the value of SD is $1.09 \times 10^{-2}$ and $5.36 \times 10^{-2}$, respectively. 
Although these functional behaviors of $c$ with respect to $\alpha$ are different between the isotropic and anisotropic cases, we expect that the anisotropic moduli space captures the tendency of the string landscape. 
The origin of this functional behavior is unclear, 
but we conclude that $\alpha \simeq 0.5$ is preferred in the SUSY AdS landscape. 
Indeed, our numerical results exhibit that a strong version of AdS distance conjecture (\ref{eq:AdS distance conjecture}) with $\alpha =1/2$ holds at the SUSY AdS vacua, independent of fluxes and VEVs. 
These results support the validity of the AdS distance conjecture in addition to the previously known Type IIB supergravity solution. 
 
For the isotropic case, we can find an upper bound of $c$ analytically which is relevant to our numerical results. Indeed, with an assumption $\mathrm{Im}\tau > 1$ and taking $\epsilon = |x|^2 - 24$ in Eq. (\ref{eq:xsquared}), the upper bound of $c^2 = m^2_{\rm tower}/|\Lambda|^{2\alpha}$ is determined as
\begin{align}
    c^2 \leq \frac{8}{9} \frac{16}{3}^{1/3} \frac{\epsilon}{N_{\rm flux}} |\Lambda|^{1-2\alpha},
\end{align}
where the numerical factor has its value $\sim 1.55$. Since our discussion in section \ref{sec:AdS/moduli}
implies $|x|^2 \sim 24$ in the landscape, we assume that $\epsilon$ takes a small value and treat it as a constant. Note that $\epsilon/N_{\rm flux}$ is always positive. Then, for $|\Lambda| < 1$, which is the condition that the AdS distance conjecture originally imposed, the upper bound decreases below $\alpha = 1/2$ and increases above $\alpha = 1/2$. Although the SD and the upper bound are different quantities, the upper bound can be expected to capture the behavior of the SD due to the condition $c^2 > 0$. In fact, the behavior of the upper bound with respect to $\alpha$ is consistent with the numerical results. 
In particular, the upper bound does not vanish under the strict limit of $\Lambda \rightarrow 0$ for the $\alpha = 1/2$ case. This result is consistent with the strong version of the AdS distance conjecture. Moreover, the upper bound is controlled by the $\mathcal{O}(1)$ tadpole charge $N_{\rm flux}$ in that case. Thus the AdS distance conjecture is also correlated with the tadpole charge or the tadpole cancellation condition. 
\begin{figure}[H]
\begin{minipage}{1.0\hsize}
\centering
     \includegraphics[page=4, width=150mm]{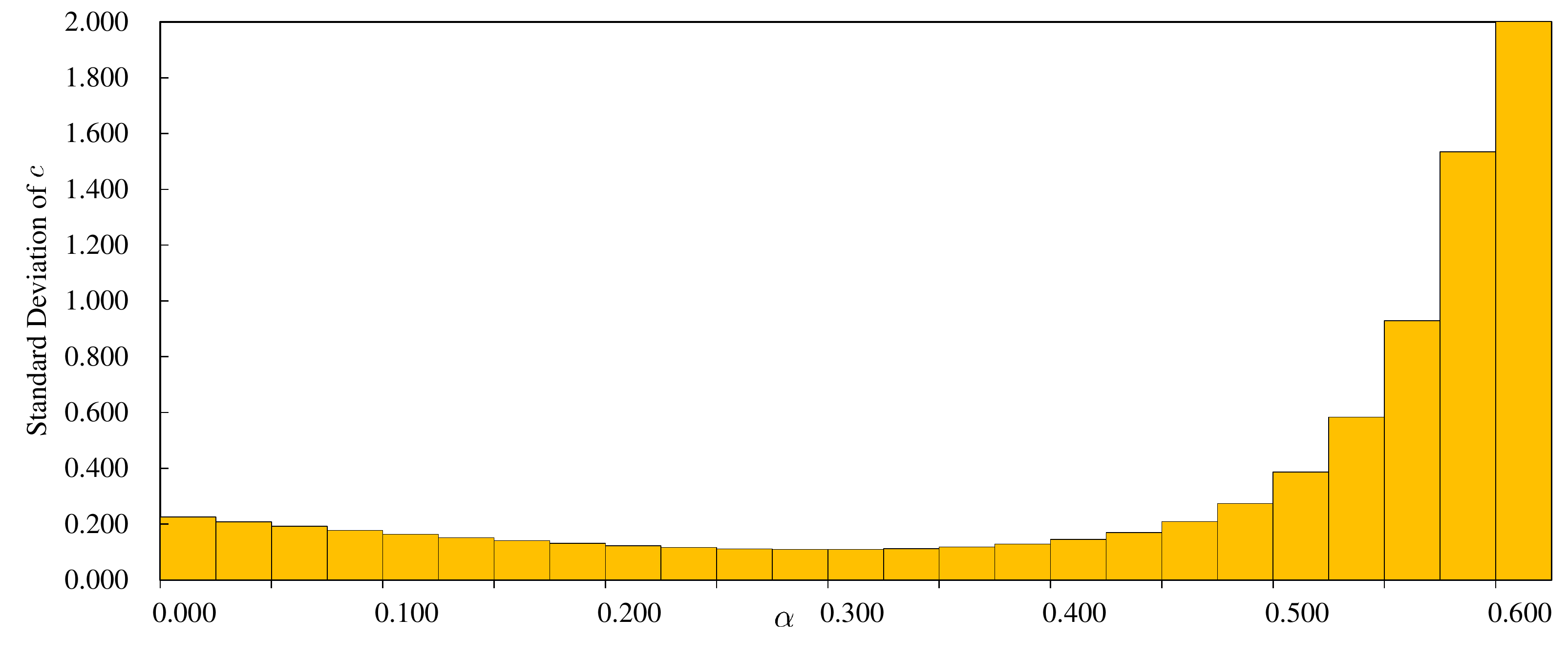}
\end{minipage}
    \caption{The standard deviation of $c = m_{{\rm tower}}/{|\Lambda|^\alpha}$ for each $\alpha$ in the case of isotropic non-SUSY vacua. The range of $\alpha$ shown in the figure is set as $[0.000, 0.600]$ and the increments are 0.025. In this figure, $\alpha = 0.275$ gives the minimum value $1.18 \times 10^{-2}$.}
    \label{fig:SD_isotropic_nonSUSY}
\end{figure}

\begin{figure}[H]
\begin{minipage}{1.0\hsize}
\centering
     \includegraphics[page=2, width=150mm]{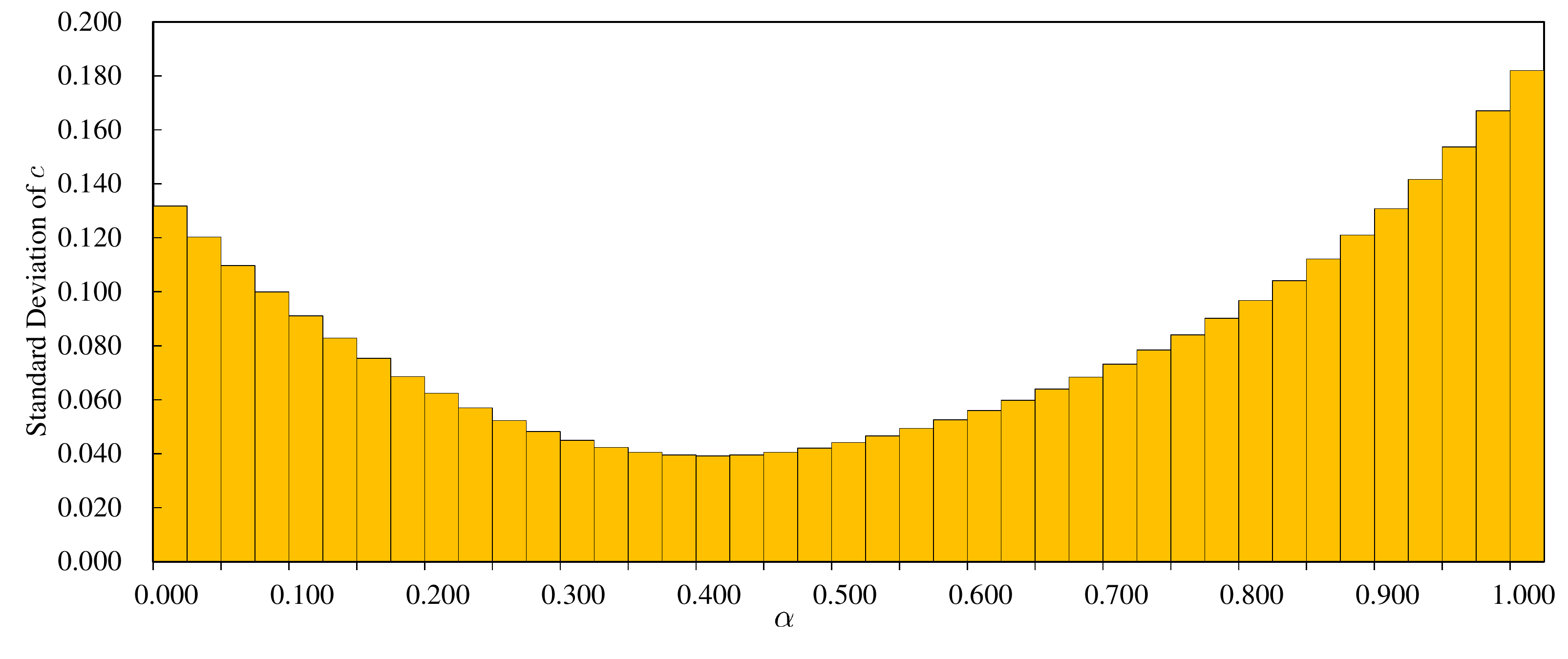}
\end{minipage}
    \caption{The standard deviation of $c = m_{{\rm tower}}/{|\Lambda|^\alpha}$ for each $\alpha$ in the case of anisotropic non-SUSY vacua. The range of $\alpha$ shown in the figure is set as $[0.000, 1.000]$ and the increments are 0.025. In this figure, $\alpha = 0.400$ gives the minimum value $3.92 \times 10^{-2}$.}
    \label{fig:SD_anisotropic_nonSUSY}
\end{figure}

This analysis can also be applied to the stable non-SUSY AdS vacua shown in Tables  \ref{tab:Numberofvacua_isotropic_orbifold} and \ref{tab:Numberofvacua_anisotropic_orbifold}, although there is no guidance to determine both $c$ and $\alpha$. 
After a statistical calculation, 
it is found that the same pattern also appears for the non-SUSY cases, as shown in Figs.  \ref{fig:SD_isotropic_nonSUSY} and \ref{fig:SD_anisotropic_nonSUSY}. 
In particular, in the anisotropic case of Fig. \ref{fig:SD_anisotropic_nonSUSY}, the minimum exists around $\alpha=0.400$ with the value of SD being $3.92 \times 10^{-2}$. 
We emphasize that these characteristic features exist in both SUSY and non-SUSY AdS vacua, but it is difficult to find out the origin of these functional behaviors in the non-SUSY case. 
We leave us to clarify this point in the future.

\subsection{The dS swampland conjecture}
\label{sec:dS}
At last, we consider the dS swampland conjecture proposed in Refs. \cite{Obied:2018sgi,Andriot:2018wzk,Garg:2018reu,Ooguri:2018wrx}, stating that no (meta-)stable dS vacuum can be obtained in consistent string compactifications\footnote{See for the recent discussion of tachyonic de Sitter solutions of 10d type II supergravities \cite{Andriot:2021rdy}.}. 
The original version of the conjecture suggests that the scalar potential $V$ in units of the reduced Planck mass $M_{\rm Pl} = 1$ satisfies
\begin{align}
    |\nabla V| \geq c \cdot V \label{eq:the original dS conjecture}  
\end{align}
with $c$ being a positive $\mathcal{O}(1)$ constant and $|\nabla V|$ denoting the norm of the gradient of $V$. Hence, no stable dS vacuum is expected to exist in the landscape. Indeed, we could not find any stable dS vacuum in the landscape with $N_{\rm flux} \leq 12$ as we showed by our numerical search in section \ref{sec:distributions}.

However, what we found in the discussion in section \ref{sec:signV} is that $N_{\rm flux} > 0$ is required to obtain stable dS or Minkowski vacua. In other words, there remains possibility that stable dS vacua exist beyond the tadpole cancellation condition: $N_{\rm flux} > 12$.
We did not explore such a region in our previous numerical search in section \ref{sec:distributions} since we focused on the landscape. 
Therefore, in this section, we extend the search region to find stable dS/Minkowski vacua by focusing on the isotropic moduli whose K\"ahler potential $K$ and superpotential $W$ are given by Eqs. (\ref{eq:Kahlerpotential_isotropic_orbifold}) and (\ref{eq:superpotential_isotropic_orbifold}), respectively. The flux quanta are randomly generated within
\begin{align}
  -20 \leq  \{f^0, f^1, f_0, f_1, h^0, h^1, h_0, h_1\} \leq 20, \label{eq:the flux quanta in the isotropic search}
\end{align}
leading to
\begin{align}
  0 \leq N_{\rm flux} \le 300.
\end{align}
We should emphasize again that the tadpole cancellation condition restricts $N_{\rm flux} \leq 12$ but we searched the region $N_{\rm flux} > 12$ violating the tadpole cancellation condition to clarify what quantities characterize the existence of dS vacua.

\begin{figure}[H]
 \begin{center}
     \includegraphics[page=1, width=160mm]{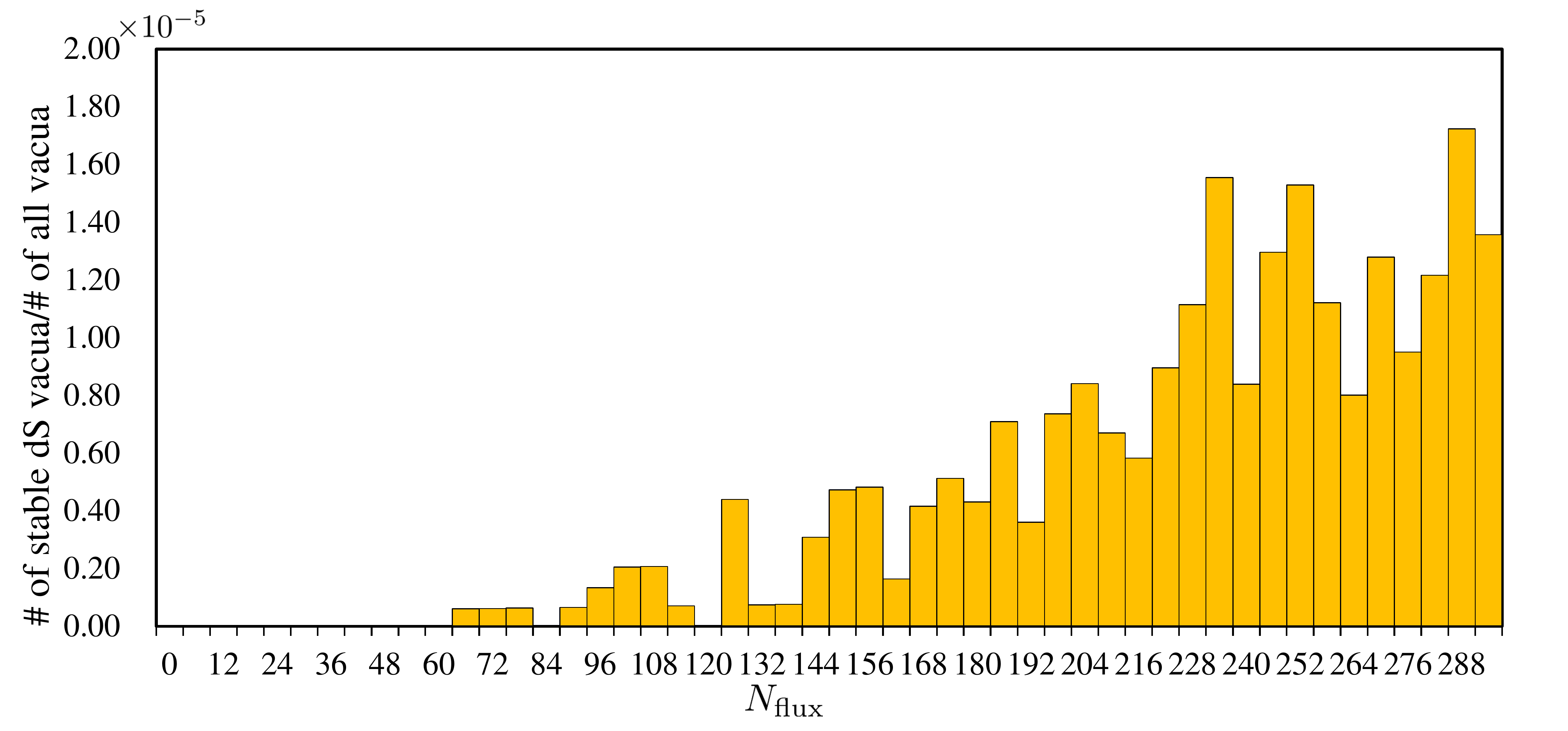}
 \end{center}
    \caption{The ratio of the number of stable dS vacua and that of all vacua against $N_{\rm flux}$. }
    \label{fig:dSvsNflux}
\end{figure}

When we enlarge the range of $N_{\rm flux}$, 
we found 238 stable dS vacua and 1 stable Minkowski vacua in the region with $N_{\rm flux} > 12$. The whole number of independent flux patterns and vacua containing unstable ones are $785{,}273{,}879$ and $66{,}751{,}429$, respectively. All of the dS vacua we found in the region $N_{\rm flux} \leq 12$ are perturbatively unstable. We list the benchmark points in Table \ref{tab:The benchmark points of Minkowski and dS vacua}. The dS and Minkowski vacua are still rare, but it is easier to obtain them outside the string landscape, namely violating the tadpole cancellation condition.  
Indeed, we plot the ratio of the number of stable dS vacua to that of all against $N_{\rm flux}$ in Fig. \ref{fig:dSvsNflux} to discuss the mutual relation between the existence of dS vacua and $N_{\rm flux}$. 
It turns out that the stable vacua begin to appear when $N_{\rm flux}$ exceeds 12, which is the maximum value of $N_{\rm flux}$ allowed by the O-planes. 
The number of the vacua increases as $N_{\rm flux}$ increases. It indicates that $N_{\rm flux}$ would be the quantity that characterizes the boundary between the landscape and the swampland.\footnote{Moreover, we observed that the number of stable AdS vacua decreases as $N_{\rm flux}$ increases in the $N_{\rm flux} > 12$ region.} 

To make the structure that $N_{\rm flux}$ characterizes the boundary between the landscape and the swampland be more apparent, let us note that there is the refined version of the dS swampland conjecture \cite{Ooguri:2018wrx}. It states that the scalar potential $V$ satisfies either Eq. (\ref{eq:the original dS conjecture}) or
\begin{align}
     \operatorname{Min} (\nabla_i \nabla_j V) \leq - c' \cdot V, \label{eq:the refined dS conjecture}
\end{align}
where $M_{\rm Pl} = 1$ and $c'$ is a positive $\mathcal{O}(1)$ constant. $\operatorname{Min} (\nabla_i \nabla_j V)$ denotes the minimum eigenvalue of the squared physical masses. The difference from the original version is that this expression allows the existence of unstable dS vacua. Indeed, we observed that the unstable dS vacua commonly exist in the region $N_{\rm flux} > 12$.
Moreover, we analyzed the region $0 \leq N_{\rm flux} \leq 12$ intensively, which is part of the landscape that we searched in section \ref{sec:distributions}. We set the flux quanta as in Eq. (\ref{eq:the flux quanta in the isotropic search}) again and prepare flux patterns $5.41 \times 10^7$ leading to the $6.90 \times 10^6$ the number of vacua. 
As a result, we found three unstable dS vacua in the landscape. We refer to one of them as Vacuum 5 in Table \ref{tab:The benchmark points of Minkowski and dS vacua}. This fact tells us that there exist unstable dS vacua in the landscape, and the $N_{\rm flux} $ bounds the existence of (meta-)stable dS vacua rather than the unstable dS vacua. 
Furthermore, we checked the refined dS conjecture explicitly by using the unstable dS vacua. The ratio $|\operatorname{Min} (\nabla_i \nabla_j V)/V|$ is defined as an upper bound of $c'$ and our results are summarized in Table \ref{tab:dSO(1)constant}. As a result, those three vacua support the refined dS conjecture, which predict $c'$ to have $\mathcal{O}(1)$ value. 
Although the number of unstable dS vacua we obtained is small, we numerically find that the perturbatively unstable dS vacua are allowed in the landscape with correct $\mathcal{O}(1)$ value $c'$. However, stable dS vacua are only allowed in the region with $N_{\rm flux} > 12$, i.e., the swampland. 
We state in section \ref{sec:vacua structure} that $N_{\rm flux}$ plays an important role in distinguishing the landscape from the swampland in a vague way, 
These numerical results clarify the vague statement of section \ref{sec:vacua structure} that $N_{\rm flux}$ characterizes and sharpens the boundaries between the landscape and the swampland. 
This phenomenon must be relevant to the emergence of $N_{\rm flux}$ in the scalar potential (\ref{eq:The scalar potential in terms of Nflux}), but a more careful investigation of this issue is left for future work.

\begin{table}[]
\centering
\begin{tabular}{cccccc}\hline
Properties                         & Vacuum 1 & Vacuum 2 & Vacuum 3 & Vacuum 4 & Vacuum 5 \\\hline
Stable?                            & Yes       & Yes       & Yes       & Yes & No          \\
Type                               & Minkowski & dS        & dS        & dS  & dS         \\
$f^0$                              & -18       & 6         & 0       & 3     & 2          \\
$f^1$                              & -7        & -12       & -7       & 11   & -2             \\
$f_0$                              & 7         & 11        & -4        & 13  & 1                \\
$f_1$                              & -15       & -19       & 1        & 17   & 3           \\
$h^0$                              & 0         & -2        & -1         & 2  & 2             \\
$h^1$                              & 1         & -4        & 2         & -2  & -2            \\
$h_0$                              & -7        & 18        & -11        & 9  & 6             \\
$h_1$                              & 1         & -8        & -4         & 4  & 3             \\
$\langle {\rm Re} S \rangle $      & 2.000     & 1.325     & -1.349     & 1.742 & 0.967     \\
$\langle {\rm Im} S \rangle $      & 8.660     & 8.252     & 4.188     & -0.158 & 1.078         \\
$\langle {\rm Re} \tau \rangle $   & 0.500     & 0.988     & -0.538     & 5.767 & -1.247        \\
$\langle {\rm Im} \tau \rangle $   & 1.443     & 1.286     & 2.104     & 1.200  & 1.382    \\
$V$                                & 0.000     & 0.000368  & 0.0000375   & 0.000928 & 0.0306       \\
$N_{\rm flux}$                     & 150       & 190       & 74        & 235    & 10        \\\hline
\end{tabular}
\caption{The benchmark points for Minkowski and dS vacua with the isotropic complex structure modulus.}
\label{tab:The benchmark points of Minkowski and dS vacua}
\end{table}

\begin{table}[H]
\centering
\begin{tabular}{cccccc}\hline
Properties                         & Unstable dS 1 (Vac. 5) & Unstable dS 2 & Unstable dS 3 \\\hline
$f^0$                              & 2        & 1        & 4       \\
$f^1$                              & -2         & 1         & 8        \\

$f_0$                              & 1         & -7        & -17         \\
$f_1$                              & 3        & 2         & 7         \\

$h^0$                              & 2         & 2         & 1          \\
$h^1$                              & -2        & 1         & 1       \\

$h_0$                              & 6        & 0        & -2         \\
$h_1$                              & 3         & 1        & 1         \\

$V$                                & 0.0306   & 8.62 $\times 10^{-4}$   & 1.72 $\times 10^{-4}$  \\
$N_{\rm flux}$                     & 10        & 11        & 12      \\
$c'$  ($\mathcal{O}(1)$ parameter)      & 5.03      & 13.30       & 6.96 \\
\hline
\end{tabular}
\caption{The $\mathcal{O}(1)$ parameter of the refined dS conjecture at  three unstable dS vacua.}
\label{tab:dSO(1)constant}
\end{table}

\section{Conclusions and Discussions}
\label{sec:con}

We extensively studied the vacuum structure of 4D effective 
field theories arising from Type IIB flux compactifications 
on the mirror of the rigid CY threefold. 
Since all the closed string moduli can be stabilized by three-form 
fluxes themselves due to the absence of K\"ahler structure 
deformations, such a class of flux compactifications 
plays a crucial role in revealing the vacuum structure of flux vacua 
and in testing the swampland conjectures. 
Remarkably, one can deal with more general fluxes than the 
T-dual Type IIA flux compactification, namely the DGKT model \cite{DeWolfe:2005uu}.

It turned out that the supersymmetric and non-supersymmetric 
AdS vacua are perturbatively stable for a large region of flux 
quanta, and there exist tachyonic directions for dS vacua in the string landscape. 
It motivated us to explore what determines the boundaries between these stable and unstable flux vacua. 
In particular, we thoroughly investigated the relationship between 
the tadpole charge and the existence of dS/Minkowski vacua. 
As analytically discussed in section \ref{sec:vacua structure}, the tadpole charge is severely constrained by the stability condition of the moduli fields in addition to the tadpole cancellation condition. 
Indeed, our numerical results exhibit that the stable dS/Minkowski vacua are only allowed in the swampland, where 
the tadpole cancellation condition is violated due to the limited range of flux quanta. 
Hence, the boundaries between the string landscape and the swampland would be 
determined by the tadpole charge. 
Since we relied on the numerical search to find such a correlation, it would be 
interesting to reveal the underlying structure and 
confirm this relation in the broader class of string compactifications, 
which we leave for future work. 

We also analyzed the AdS/moduli scale separation and the AdS distance conjectures. Our analytical and numerical results exhibit that 
${\cal O}(1)$ parameters in expressions of AdS/moduli scale separation conjecture peaked around specific values in both supersymmetric and non-supersymmetric compactifications. 
Such sharp distributions are peculiar to the landscape since only a small tadpole charge is consistent with the tadpole cancellation condition. 
Moreover, the standard deviation of ${\cal O}(1)$ parameter 
in the expression of AdS distance conjecture is minimized around $\alpha \simeq 0.5$ 
with respect to a power of cosmological constant $|\Lambda|^\alpha$, 
although there exists a slight difference between supersymmetric and non-supersymmetric compactifications. 
It is remarkable that these phenomena are strongly correlated with the tadpole charge, which determines the structure of AdS vacua. 
In this way, our results show the hidden structure of the string landscape. 
It would be interesting to figure out their 
origin in other corners of the string landscape.

\subsection*{Acknowledgments}

We would like to thank H. Abe and S. Mizoguchi for useful discussions and comments. 
H. O. was supported in part by JSPS KAKENHI Grant Numbers 
JP19J00664 and JP20K14477.

\appendix

\section{Notation of the special geometry}
\label{sec:Special Geometry}
In this section, we summarize the notation of the special geometry in this paper, following Ref. \cite{Candelas:1990pi} (it was also reviewed in Ref. \cite{Becker:2007ee}).

First, let us introduce the harmonic basis of $H^3(\mathscr{M}, \mathbb{C})$:$\{\Omega,\chi_{a},\bar{\chi}_{\bar{b}},\Omega\}$ with $a,b = 1,\dots,h^{2,1}$. 
As mentioned in Eq. (\ref{eq:derivative of Omega by complex-structure}), $\Omega$ and $\chi_i$ are connected by a certain differentiation. 
Since $\Omega$ is defined up to a holomorphic multiplication
\begin{align}
  \Omega \rightarrow f(u) \Omega,
\end{align}
the usual differentiation should be replaced with a covariant derivative $D$. 
This transformation is often called the K\"ahler transformation. More concretely, $\Omega$ is a holomorphic section of $\mathcal{H} \otimes L$, where $\mathcal{H}$ and $L$ denote a $Sp(2n+2, \mathbb{R})$ vector bundle and a line bundle, respectively. $\mathcal{H}$ is attributed to the definition of the real basis of $H^3(\mathscr{M}, \mathbb{C})$ given in Eq. (\ref{eq:dual cohomology basis}). The K\"ahler covariant derivative is defined to be covariant on $L$. 
In general, we can consider a quantity $\psi$ that transforms under the K\"ahler transformation as
\begin{align}
 \psi^{(m, n)} \rightarrow f^m \bar{f}^n \psi^{(m, n)},
\end{align}
i.e., $\psi^{(m, n)} \in L^m \otimes \bar{L}^{\bar{n}}$ (we ignored a choice of the basis).
Then the covariant derivative is generalized to act on $\psi$ as
\begin{align}
D_a \psi^{(m, n)} &= (\partial_a +  m K_a)  \psi^{(m, n)}, \\
D_{\bar{b}} \psi^{(m, n)} &= (\partial_{\bar{b}} +  n K_{\bar{b}})  \psi^{(m, n)},
\end{align}
so that these transformations are the same as the original $\psi^{(m, n)}$, 
\begin{alignat}{2}
  D_a \psi^{(m, n)} &\rightarrow f^{m} \bar{f}^{n} D_a \psi^{(m, n)}, & \quad  D_{\bar{b}} \psi^{(m, n)} &\rightarrow f^{m} \bar{f}^{n} D_{\bar{b}} \psi^{(m, n)}.
\end{alignat}
When $\psi^{(m, n)}$ is a tensor, we have to use $\nabla_a, \nabla_{\bar{b}}$ instead of $\partial_a, \partial_{\bar{b}}$ with Christoffel symbols $\Gamma^{a}_{b c}, \Gamma^{\bar{a}}_{\bar{b} \bar{c}}$ defined on a K\"ahler manifold.
We recall that the K\"ahler metric is a constant under the covariant derivative, i.e., $D_c g_{a \bar{b}} = 0$. 

Let us summarize how these covariant derivatives act on the basis $\{\Omega, \chi_a, \bar{\chi}_{\bar{b}}, \bar{\Omega}\}$. First of all, 
\begin{align}
  D_a \Omega = \chi_a
\end{align}
already follows from Eq. (\ref{eq:derivative of Omega by complex-structure}), and $D_a \bar{\Omega} = 0$ holds for a similar reason. 
Before proceeding to analyze the transformations $\{\chi_a, \bar{\chi}_{\bar{b}}\}$, we introduce the Yukawa coupling\cite{Strominger:1985it, Candelas:1990pi} \footnote{There is a subtle difference between this definition of Yukawa coupling and that in Ref. \cite{Candelas:1990pi} that here $a, b, c$ are indices of $\{z^a\}$, but not of $\{u^a\}$. However, if we discuss only on the $z^0 = 0$ patch this makes no difference.}
\begin{align}
  \kappa_{abc} &= + i \langle \Omega, D_{a} D_{b} D_{c} \Omega \rangle = + \int_{\mathscr{M}} \Omega \wedge D_{a} D_{b} D_{c} \Omega,
\end{align}
where the inner product $\langle \alpha, \beta \rangle$ is defined as
\begin{align}
  \langle \alpha, \beta \rangle &= -i \int_{\mathscr{M}} \alpha \wedge \beta.
\end{align}
The expression of $\kappa_{abc}$ reduces to 
\begin{align}
  \kappa_{abc} &= i \langle \Omega, \partial_{a} \partial_{b} \partial_{c} \Omega \rangle,
\end{align}
since only the $(0, 3)$- part of $D_{a} D_{b} D_{c} \Omega$ contributes to the integration, and $\kappa_{a b c}$ is thus totally symmetric with respect to its indices.\footnote{These Yukawa couplings are of phenomenological interest in the context of heterotic string with standard embedding as recently discussed in Ref.\cite{Ishiguro:2021drk}.}

In our notation, $ \left\langle \Omega \wedge \bar{\Omega} \right\rangle$ is equal to 
\begin{align}
  \langle \Omega \wedge \bar{\Omega} \rangle  = + e^{-K_{\rm cs}}.
\end{align}
This yields another expression of the K\"ahler metric $g_{a \bar{b}}$ 
\begin{align}
  \langle \chi_a, \bar{\chi}_{\bar{b}} \rangle = - g_{a \bar{b}} e^{-K_{\rm cs}}.
\end{align}
Although 
\begin{align}
  [D_a, D_{\bar{b}}] \Omega &= -g_{a \bar{b}} \Omega \label{eq:commutator of Kahler derivative on Omega}
\end{align}
follows from the above results, but this just reflects the fact that
\begin{align}
  c_1 (L) &= [J],
\end{align}
with $J$ being the K\"ahler form.

Following the lines of Ref. \cite{Candelas:1990pi}, it follows that
\begin{align}
  D_{a} \chi_b &= -i e^{K_{\rm cs}} {\kappa_{a b}}^{\bar{c}} \bar{\chi}_{\bar{c}}.
\end{align}
By using Eq. (\ref{eq:commutator of Kahler derivative on Omega}) and $D_{\bar{a}} \Omega$, we also obtain
\begin{align}
  D_{a} \bar{\chi}_{\bar{b}} &= g_{a \bar{b}} \bar{\Omega}.
\end{align}

\section{Calculation of the scalar potential}
\label{app:cal}

The detailed calculation of the scalar potential is given in this section with an emphasis on the $k_S = 4$ case. The results, which would be given below, hold for the $k_S = 1$ case similarly. 
The K\"ahler potential we consider is given by
\begin{align}
  K = -4 \log\left(-i \left(S-\bar{S}\right)\right) -  \log\left(-i \int \Omega \wedge \bar{\Omega} \right). 
\end{align}
Let us split $\tilde{V}$ into the following three pieces under the assumption that the moduli spaces of the axio-dilaton and complex structure are product;
\begin{align}
  \tilde{V}_S &= K^{S\bar{S}} D_S W D_{\bar{S}} \bar{W},\\ 
  \tilde{V}_{\rm cs} &= \sum_{a, b} K^{a\bar{b}}D_a W D_{\bar{b}} \bar{W},\\
  \tilde{V}_{\rm SG} &= - 3 |W|^2,
\end{align}
where $a, b$ run over the complex structure moduli. 
In the following, we will derive the quantity $\partial_{{\rm Im}S}\tilde{V}$ whose sign is directly related to the sign of the cosmological constant. 
The derivation is performed by calculating three pieces: $\partial_{{\rm Im}S}\tilde{V}_S$, $\partial_{{\rm Im}S}\tilde{V}_{\rm cs}$ and $\partial_{{\rm Im}S}\tilde{V}_{\rm SG}$. 
\begin{itemize}
    \item $\partial_{{\rm Im}S}\tilde{V}_S$
    
    First, we calculate $\partial_{{\rm Im}S}\tilde{V}_S$ whose explicit form is given by
\begin{align}
  \partial_{{\rm Im}S}\tilde{V}_S = 2 |W_{\rm NS}|^2 {\rm Im}S + 4 {\rm Im} (W_{\rm NS} \bar{W}_{\rm RR}),
\end{align}
with $W_{\rm RR} = \int F_3 \wedge \Omega$, $W_{\rm NS} = \int H_3 \wedge \Omega$ and $W = W_{\rm RR} - S W_{\rm NS}$. 

    \item $\partial_{{\rm Im}S}\tilde{V}_{\rm cs}$
    
Next, we consider $\partial_{{\rm Im}S}\tilde{V}_{\rm cs}$. 
Note that we do not assume that the K\"ahler metric of complex structure moduli $K_{a\bar{b}}$ is diagonal.  
We recall that $K_{a\bar{b}}$ can be expressed as
\begin{align}
  K_{a \bar{b}} = - \frac{\int{\chi_a \wedge \bar{\chi}_{\bar{b}}}}{\int{\Omega \wedge \Omega}},
  \label{eq:metric for complex-structure moduli space}
\end{align}
and 
\begin{align}
  D_a W = \int G_3 \wedge \chi_a
\end{align}
from the fact that $k_a = - K_a = - \partial_a K$ with $k_a$ in Eq. (\ref{eq:derivative of Omega by complex-structure}) holds. 
Hence $\tilde{V}_{\rm cs}$ takes the form
\begin{align}
  \tilde{V}_{\rm cs} = \sum_{a,b} K^{a\bar{b}} \int{G_3 \wedge \chi_a} \int{\bar{G_3} \wedge \bar{\chi}_{\bar{b}}},
\end{align}
and then the derivative of $\tilde{V}_{\rm cs}$ becomes
\begin{align}
  \partial_{{\rm Im}S} \tilde{V}_{\rm cs} &= 2 {\rm Im} S \sum_{a, b} K^{a\bar{b}}\int{H_3 \wedge \chi_a} \int{H_3 \wedge \chi_{\bar{b}}}  \nonumber \\ &+ i \sum_{a, b} K^{a\bar{b}} \left( \int{F_3 \wedge \chi_a} \int{H_3 \wedge \bar{\chi}_{\bar{b}}} -  \int{H_3 \wedge \chi_a} \int{F_3 \wedge \bar{\chi}_{\bar{b}}} \right).
\end{align}
If we expand real three-forms $\{F_3, H_3\}$ in the basis $\{\Omega, \chi_I, \bar{\chi}_{\bar{J}}, \bar{\Omega}\}$ of $H^3(\mathscr{M}, \mathbb{C})$ as
\begin{align}
\begin{split}
  F_3 &= A_F \Omega + \sum_{a} B_{F, a} \chi_a + \sum_{a} \bar{B}_{F, \bar{a}} \bar{\chi}_{\bar{a}} + \bar{A}_{F} \bar{\Omega} 
  \\
  H_3 &= A_H \Omega + \sum_{a} B_{H, a} \chi_a + \sum_{a} \bar{B}_{H, \bar{a}} \bar{\chi}_{\bar{a}} + \bar{A}_{H} \bar{\Omega}, \label{eq:expansionFH}
\end{split}
\end{align}
the zero-th order term in $\partial_{{\rm Im}S} \tilde{V}_{\rm cs}$ is provided by
\begin{align}
  \partial_{{\rm Im}S} \tilde{V}_{\rm cs} 
  &\supset -iK^{a\bar{b}} \sum_{c, d}\left(\bar{B}_{F, \bar{c}} B_{H, d} - B_{F, d} \bar{B}_{H, \bar{c}} \right) \int{\chi_a \wedge \bar{\chi}_{\bar{c}}} \int \chi_d \wedge \bar{\chi}_{\bar{b}} \nonumber \\
  &= -iK^{a\bar{b}} \sum_{c, d}\left(\bar{B}_{F, \bar{c}} B_{H, d} - B_{F, d} \bar{B}_{H, \bar{c}} \right) K_{a\bar{c}}K_{d\bar{b}}\left(\int{\Omega \wedge \bar{\Omega}}\right)^2 \nonumber \\
  &= -i\left(\int{\Omega \wedge \bar{\Omega}}\right)^2 \sum_{c, d} K_{\bar{c} d} \left(\bar{B}_{F, \bar{c}} B_{H, d} - B_{F, d} \bar{B}_{H, \bar{c}}\right), 
  \label{eq:first order term in Vcs}
\end{align} 
where we used Eq. (\ref{eq:metric for complex-structure moduli space}) from the first line to the second line and summed up with $\{a, \bar{b}\}$ to lead the third equality.

    \item $\partial_{{\rm Im}S}\tilde{V}_{\rm SG}$

Lastly, $\partial_{{\rm Im}S} \tilde{V}_{\rm SG}$ is obtained as
\begin{align}
  \partial_{{\rm Im}S} \tilde{V}_{\rm SG} = -6 {\rm Im}\left( W_{\rm NS} \bar{W}_{\rm RR} \right) - 6 {\rm Im}S |W_{\rm NS}|^2 .
\end{align}
\end{itemize}

As a result, we arrive at
\begin{align}
  \partial_{{\rm Im}S} \tilde{V} =& \partial_{{\rm Im}S} \tilde{V}_S +\partial_{{\rm Im}S} \tilde{V}_{\rm cs}+\partial_{{\rm Im}S} \tilde{V}_{\rm SG}
   \nonumber\\
  =&2 \left( -2|W_{\rm NS}|^2 + \sum_{a, b} K^{a\bar{b}} \int{H_3 \wedge \chi_a} \int{H_3 \wedge \bar{\chi}_{\bar{b}}}  \right) {{\rm Im} S} -2 {\rm Im} \left( W_{\rm NS} \bar{W}_{\rm RR} \right) \nonumber\\ &  - i \left(\int{\Omega \wedge \bar{\Omega}}\right)^2 \sum_{a, b} K_{\bar{a}b} \left( \bar{B}_{F, \bar{a}} B_{H, b} - B_{F, b} \bar{B}_{H, \bar{a}} \right). \label{eq:derivative of tildeV with ImS}
\end{align}
In fact, the first-order term of $\tilde{V}$ with respect to ${\rm Im}S$ is proportional to $N_{\rm flux}$. Using the expansions (\ref{eq:expansionFH}), we obtain
\begin{align}
  N_{\rm flux} &= \int{H_3 \wedge F_3}  \nonumber \\
               &= - 2 {\rm Re} \left( \bar{A}_H A_F \int{\Omega \wedge \bar{\Omega}}\right) - \left(\int{\Omega \wedge \bar{\Omega}}\right) \sum_{a, b} K_{a\bar{b}} \left( B_{H, a} \bar{B}_{F, \bar{b}} - B_{F, a} \bar{B}_{H, \bar{b}} \right). \label{eq:expansionNflux}
\end{align}
Meanwhile, by rewriting the zeroth-order term of (\ref{eq:derivative of tildeV with ImS}) into
\begin{align}
  \left.\partial_{{\rm Im}S} \tilde{V}\right|_{{\rm Im}S \rightarrow 0} =& 2 {\rm Im} \left(\int{\Omega \wedge \bar{\Omega}}\right) {\rm Re} \left( \bar{A}_H A_F \int{\Omega \wedge \bar{\Omega}}\right)  \nonumber \\&+ {\rm Im}\left( \int{\Omega \wedge \bar{\Omega}}\right) \sum_{a, b} K_{a\bar{b}} \left( B_{H, a} \bar{B}_{F, \bar{b}} - B_{F, a} \bar{B}_{H, \bar{b}} \right), 
\end{align}
with $\int{\Omega \wedge \bar{\Omega}}$ being a pure imaginary quantity we conclude that 
\begin{align}
  \left.\partial_{{\rm Im}S} \tilde{V}\right|_{{\rm Im}S \rightarrow 0} = - e^{-K_{\rm cs}} N_{\rm flux}
\end{align}
holds for the $k_S = 4$ case. It is remarkable that the same expression can also be obtained for the $k_S=1$ corresponding to the no-scale scalar potential\footnote{We define $\tilde{V} \equiv e^{-K} \tilde{V}$ again for the no-scale case, but with $ K =\log\left(-i \left(S-\bar{S}\right)\right) -  \log\left(-i \int \Omega \wedge \bar{\Omega} \right) - 3 \log \left( -i (T - \bar{T})\right)$ with $T$ being a K\"ahler modulus. Here and what in follows, the $k_S = 1$ case with this different $K$ and $V$ is called the no-scale type.}.

It turned out that $\tilde{V}$ is of the form
\begin{align}
  \tilde{V} = \frac{1}{2} \partial^2_{{\rm Im}S} \tilde{V} ({\rm Im}S)^2 - e^{-K_{\rm cs}} N_{\rm flux} {\rm Im} S + C
\end{align}
with 
\begin{align}
C = \left.|W|^2\right|_{{\rm Im} S \rightarrow 0} + \left.G^{a \bar{b}}D_a W D_{\bar{b}} \bar{W}\right|_{{\rm Im}S \rightarrow 0}.
\label{eq:C}
\end{align}

\section{Expression of the tadpole charge at SUSY vacua}
\label{app:Nflux}

In this section, we show the simplified expression of $N_{\rm flux}$ derived 
in Appendix \ref{app:cal}. 
First, we focus on the SUSY condition for the axio-dilaton direction 
\begin{align}
  D_S W &= (\partial_S + K_S) W = -W_{\rm NS} - 4 \frac{W}{S - \bar{S}} = 0,
\end{align}
where $W_{\rm NS}$ and $W_{\rm RR}$ are expressed as
\begin{alignat}{2}
  W_{\rm RR} &= \bar{A}_{F} \int \bar{\Omega} \wedge \Omega, \quad & W_{\rm NS} &= \bar{A}_{H} \int \bar{\Omega} \wedge \Omega.
\end{alignat}
Hence, it allows us to eliminate $A_F$ by $A_H$, 
\begin{align}
  A_F       &= + \frac{A_H}{4} \left(S + 3\bar{S}\right).
\end{align}
On the other hand, the SUSY conditions of the complex structure sector $D_a W = 0$ 
lead to 
\begin{align}
  0 = D_a W = K_{a \bar{b}} \left( \bar{B}_{F, \bar{b}} - S \bar{B}_{H, \bar{b}} \right) \int{\Omega \wedge \bar{\Omega}}.
\end{align}
Since $\int \Omega \wedge \bar{\Omega} \neq 0$, $B_F$ can also be eliminated by $B_H$ as
\begin{align}
  B_{F, a} &= S B_{H, a}.
\end{align} 
By substituting these results into Eq. (\ref{eq:expansionNflux}), we obtain the following expression:
\begin{align}
  N_{\rm flux} &= i e^{-K_{\rm cs}} (S - \bar{S}) \left( \frac{1}{2} |A_H|^2 - K_{a \bar{b}} B_{H, a} \bar{B}_{H, \bar{b}} \right) \nonumber \\
               &= i e^{+K_{\rm cs}} (S - \bar{S}) \left( \frac{1}{2} |W_{\rm NS}|^2 - K^{a \bar{b}} D_a W_{\rm NS} D_{\bar{b}} \bar{W}_{\rm NS} \right)
               \label{eq:Nflux on SUSY vacua}
\end{align}
at the SUSY vacua, where we used $A_H = -i e^{K_{\rm cs}} \bar{W}_{\rm NS}$ and 
\begin{align}
  \sum_{a, \bar{b}} K_{a \bar{b}} B_{H, a} \bar{B}_{H, \bar{b}} &= e^{2K_{\rm cs}} K^{a\bar{b}} \int{H_3 \wedge \chi_a} \int {H_3 \wedge \bar{\chi}_{\bar{b}}}.
\end{align}
Furthermore, when we use $D_S W = 0$ again, $N_{\rm flux}$ is correlated with the AdS scale, i.e., 
\begin{align}
  \frac{N_{\rm flux}}{\Lambda_{\rm AdS}} = \frac{1}{3} (- i (S - \bar{S}))^{3} \left( 8 - |S - \bar{S}|^2 K^{a \bar{b}} \frac{D_a W_{\rm NS} D_{\bar{b}} \bar{W_{\rm NS}}}{|W|^2} \right).
  \label{eq:NfluxAdS}
\end{align}

\bibliography{references}
\bibliographystyle{h-physrev}

\end{document}